\documentclass{PoS}
\usepackage[italian,english]{babel}
\usepackage{braket}
\usepackage{amssymb}
\usepackage{amsfonts}
\usepackage{amsmath}
\usepackage{fancyhdr}
\usepackage{lineno}
\usepackage{subfigure}
\newcommand{\nn}{\nonumber}
\newcommand{\lsim}{\mathrel{\mathop{\kern 0pt \rlap
  {\raise.2ex\hbox{$<$}}}
  \lower.9ex\hbox{\kern-.190em $\sim$}}}
\newcommand{\gsim}{\mathrel{\mathop{\kern 0pt \rlap
  {\raise.2ex\hbox{$>$}}}
  \lower.9ex\hbox{\kern-.190em $\sim$}}}

\newcommand{\be}{\begin{equation}}
\newcommand{\ee}{\end{equation}}
\newcommand{\beqa}{\begin{eqnarray}}
\newcommand{\eeqa}{\end{eqnarray}}
\newcommand{\bea}{\begin{eqnarray}}
\newcommand{\eea}{\end{eqnarray}}

\newcommand{\beq}{\begin{equation}}
\newcommand{\eeq}{\end{equation}}

\newcommand{\sm}{\mathcal{S}}

\usepackage{pstricks}
\usepackage{color}
\usepackage{multirow}
\usepackage{cite}

        \let\m=\mu
\let\n=\nu

\newcommand{\sdfrac}[2]{\mbox{\small$\displaystyle\frac{#1}{#2}$}}

\title{Dimensional Regularization of Topological Terms  in Dilaton Gravity}

\ShortTitle{Topological terms in dilaton gravity}

\author{\speaker{Claudio Corian\`o}%
\\
       Dipartimento di Matematica e Fisica "Ennio De Giorgi", \\ Universit\`a del Salento and INFN-Lecce, \\ Via Arnesano, 73100 Lecce, Italy\\

        E-mail: \email{claudio.coriano@le.infn.it}}

\author{$^{(a) (b)}$ Mario Cret\'i, $^{(a)}$Stefano Lionetti, $^{(c)}$Matteo Maria Maglio and $^{(a)}$Riccardo Tommasi \\
       $^{(a)}$ 
        Dipartimento di Matematica e Fisica "Ennio De Giorgi", \\ Universit\`a del Salento and INFN-Lecce,  Via Arnesano, 73100 Lecce, Italy\\
    $^{(b)}$ Center for Biomolecular Nanotechnologies, Istituto Italiano di Tecnologia, Via Barsanti 14, 73010 Arnesano, Lecce, Italy.
\\
$^{(c)}$  Institute for Theoretical Physics (ITP), University of Heidelberg
Philosophenweg 16, 69120 Heidelberg, Germany\\
}

\abstract{
The possibility of evading Lovelock's theorem at $d=4$, via a singular redefinition of the dimensionless coupling of the Gauss-Bonnet term, has been extensively discussed in the cosmological context. The term is added as a quadratic contribution of the curvature tensor to the Einstein-Hilbert action, originating theories of  "Einstein Gauss-Bonnet" (EGB) type. 
These studies are interlaced with those of the conformal anomaly effective action. We review some basic results concerning the structure of these actions, their conformal constraints around flat space and their relation to EGB theories. The local and nonlocal formulations of such effective actions are illustrated. This class of theories find applications in the seemingly unrelated context of topological materials, subjected to thermal and mechanical stress. 
}

\FullConference{Proceedings of the Corfu Summer Institute 2021 "Workshops on the Standard Model and beyond "\\
	
	Corfu, Greece}

\begin{document} 
\section{Introduction}
The analysis of the conformal backreaction, describing the modifications induced on a background gravitational metric by the integration of a conformal sector, could play an important role in the physics of the early universe, setting the initial conditions for its later evolution.\\
The way we will address this point is within the context of semiclassical gravity, where the loop corrections modify the metric by integrating out the quantum matter fields. The approach follows closely Sakharov's formulation of induced gravity, where all the gravitational terms, including the Einstein-Hilbert one, are induced by quantum matter propagating in a curved spacetime, where the original metric is left to "freely flap in the breeze"  \cite{Visser:2002ew}.

At a second stage, the cosmological evolution follows by adding classical matter to such a theory of gravity, which is modified by higher powers of the curvature, in the form of a classical, conserved stress energy tensor, as in any traditional cosmological approach.  \\
The structure of the induced gravity action, in principle, covers all powers of $R$ and includes dimensional constants, coming from the quantum corrections, generating terms of the form  \cite{Visser:2002ew}
\begin{equation} 
\sm_{eff}\sim \int d^4 x \sqrt{g}\left( \Lambda + c_1(g) R + c_2``R^2" \right),
\end{equation}
corresponding to a cosmological constant, the Einstein-Hilbert action and to generic $``R^2"$ terms. However, if we limit ourselves only to the conformal contributions, there are drastic simplifications in the result, since the only dimensional parameter of the theory comes from the renormalization scale.  \\
The ordinary steps to be followed in order to characterize such conformal contributions start from  the (Euclidean) functional integral, 
\begin{equation} 
\label{induced}
Z(g)=N \int D\phi e^{-S_0(g,\phi)},   \qquad Z(\delta)=1,
\end{equation}
where $g$ is the metric and $\phi$ the quantum field that is integrated out. 
Its logarithm, $\mathcal{S}(g)$, is our definition of the effective action. $S_0(g,\phi)$ is the classical action. 
As a reference for our discussion, as already mentioned, we may assume that $S_0(\phi,g)$ describes, for instance, a free scalar field $\phi$ in a generic background. The action, in  this case, is given by 
\bea
\label{phi}
S_0(g,\phi)&=&\frac{1}{2}\int\, d^dx\,\sqrt{-g}\left[g^{\mu\nu}\nabla_\mu\phi\nabla_\nu\phi-\chi\, R\,\phi^2\right],
\eea
where we have included a conformal coupling $\chi(d)=\frac{1}{4}\frac{(d-2)}{(d-1)}$, and $R$ is the scalar curvature.
 This choice of $\chi(d)$ guarantees the conformal invariance of this action in $d$ dimensions as well as introduces a term of improvement for the stress-energy tensor in the flat limit, which becomes symmetric and traceless. 
As usual $Z(g)$, in the Feynman diagrammatic expansion, will contain both connected and disconnected graphs, while $\mathcal{S}(g)$ collects only connected graphs. It is easy to verify that this 
collection corresponds also to 1PI (1 particle irreducible) graphs only in the case of free field theories embedded in external (classical) gravity.  \\

It is summarized by the following picture, where we sum over the external graviton lines 
\begin{align}
\label{figg}
\sm(g)=& \sum_n \quad\raisebox{-8.5ex}{{\includegraphics[width=0.19\linewidth]{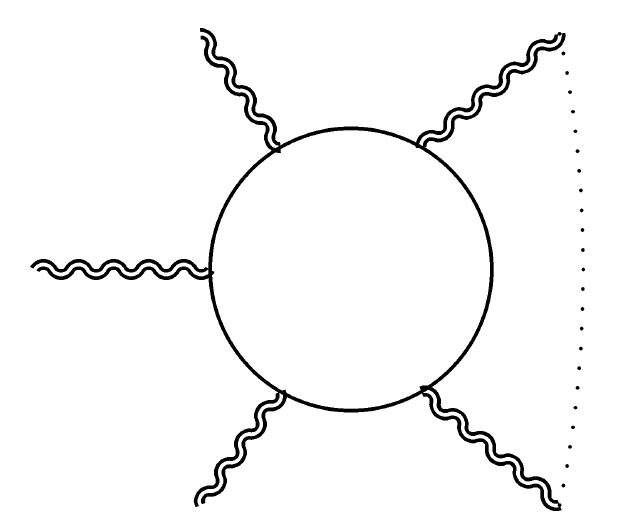}}}, \,\scriptstyle \text{(n-point)}
\end{align}
that we expand in the form
\begin{equation}
\label{exps2}
\sm(g)_B\equiv\sm(\bar{g})_B+\sum_{n=1}^\infty \frac{1}{2^n n!} \int d^d x_1\ldots d^d x_n \sqrt{g_1}\ldots \sqrt{g_n}\,\langle T^{\mu_1\nu_1}\ldots \,T^{\mu_n\nu_n}\rangle_{\bar{g} B}\delta g_{\mu_1\nu_1}(x_1)\ldots \delta g_{\mu_n\nu_n}(x_n),
\end{equation}
in terms of bare $(B)$ $nT$ correlators, with 
\begin{equation}
\label{exps1}
\langle T^{\mu_1\nu_1}(x_1)\ldots T^{\mu_n\nu_n}(x_n)\rangle_B \equiv\frac{2}{\sqrt{g_1}}\ldots \frac{2}{\sqrt{g_n}}\frac{\delta^n \sm_B(g)}{\delta g_{\mu_1\nu_1}(x_1)\delta g_{\mu_2\nu_2}(x_2)\ldots \delta g_{\mu_n\nu_n}(x_n)}, 
\end{equation}
where $\sqrt{g_1}\equiv \sqrt{|\textrm{det} \, g_{{\mu_1 \nu_1}}(x_1)} $ and so on.
The 
sum of diagrams provides an exact one-loop result, since the loop contribution is conformal if we limit ourselves to free field theory, as is here the case.  As we move down to $d=4$, the series needs to be renormalized and the special property of the result is that such renormalization involves only two specific counterterms, of deep significance: the Euler-Poincar\`e density (Gauss-Bonnet term) and the squared of the Weyl tensor.\\
\section{The counterterms}
In DR the divergences in the effective action appear as single poles in $\epsilon=d-4$ if we couple a conformal sector to gravity, and the expansion around $d=4$, for a curved background, requires some special care. "Extra" degrees of freedom of the higher dimensional metric survive on the 4-dimensional manifold, after we perform the limit. Most of the ambiguities associated with the structure of the effective action are related with this aspect of the regularization, which does not have a unique solution. We may resort to phenomenological intuition in order to appreciate the expressions of the dilaton gravities that result from the procedure. We are going to summarize some of the points investigated in the near past, underlying some key open issues that will be discussed in a forthcoming work. 
 
The two counterterms to be included are $V_E$ and $V_{C^2}$, giving a regularized effective action of the form 
\begin{equation}
{Z}_R(g)_=\, \mathcal{N}\int D\Phi e^{-S_0(g,\Phi) + b' \frac{1}{\epsilon}V_E(g,d) + b \frac{1}{\epsilon}V_{C^2}(g,d)},
\end{equation} 
where $\mathcal{N}$ is a normalization constant, and 
\begin{align}
\label{ffr}
V_{C^2}(g, d)\equiv & \mu^{\varepsilon}\int\,d^dx\,\sqrt{-g}\, C^2, \notag \\
V_{E}(g,d)\equiv &\mu^{\varepsilon} \int\,d^dx\,\sqrt{-g}\,E , 
\end{align}
are the Gauss-Bonnet  and the Weyl terms respectively, 
where $\mu$ is a renormalization scale.
The two counterterms are given in terms of the dimension-4 curvature invariants
\begin{align}
\label{fourd}
E_4&\equiv R_{\mu\nu\alpha\beta}R^{\mu\nu\alpha\beta}-4R_{\mu\nu}R^{\mu\nu}+R^2,  \\
C^{(d) \alpha\beta\gamma\delta}C^{(d)}_{\alpha\beta\gamma\delta}
&=
R^{\alpha\beta\gamma\delta}R_{\alpha\beta\gamma\delta} -\frac{4}{d-2}R^{\alpha\beta}R_{\alpha\beta}+\frac{2}{(d-2)(d-1)}R^2.\label{Geometry1}
\end{align}
Much of the discussion about the structure of  the effective action is related to the correct computation of such counterterms in dimensional Regularization (DR) and the possibility of performing a {\em further} finite renormalization of the regulated theory, in order to derive two different formulations of the effective action. The two formulations are not equivalent, since they correspond either to a quartic or to a quadratic dilaton gravity. In the case of a quadratic theory, the dilaton can be removed from the spectrum, thereby generating a nonlocal action, corresponding to the celebrated Riegert's action \cite{1984PhLB..134...56R}. The nonlocal theory is entirely formulated in terms of the original metric $g$, while quartic dilaton gravities are expressed in terms of a fiducial metric $\bar{g}$ and of a dilaton, according to the conformal decomposition 
\beq
g_{\mu\nu}=\bar{g}_{\mu\nu}e^{2 \phi}.
\label{cf1}
\eeq
In the quadratic case, as already mentioned, the dilaton can be entirely removed. 
Notice that the regularization process identifies a specific fiducial metric $\bar{g}$, which is uniquely singled out, modulo the ordinary diffeomorphisms of the background manifold. The local shift symmetry  
\beq
g_{\mu\nu}\to g_{\mu\nu}e^{-2 \sigma},\qquad \phi\to \phi + \sigma
\label{cf2}
\eeq
which is part of the conformal decomposition above, is broken as we try to remove the divergences of the effective action. This is quite obvious if one looks at the structure of the Wess-Zumino form of the effective action, which is expressed as difference of a functional in the $g$ and $\bar{g}$ metrics. In other words, the fundamental fields of the renormalized theory are the fiducial metric {\em and} the dilaton. The dynamics, as we are going to review,  is constrained by the anomaly that relates the equations of motion of $\phi$ and $\bar{g}$. \\
 This reformulation of the quartic effective action is also accompanied by the appearance of a new fundamental scale $f$, characterising the breaking of the conformal symmetry. The scale is simply introduced by normalizing the dilaton to be of mass dimension one: $\phi\to \phi/f$.
 For this reason, one may wonder about the significance of both actions, which may cover completely different energy regions in the dynamics of the early universe.
While local actions can be expanded in the $1/f$ scale - up to quartic order - in the  nonlocal formulation, the expansion variable is $R\Box^{-1}$, which is dimensionless and suitable for a description of the effective action close to the Planck scale. This expansion has been compared, around flat spacetime, with the prediction coming from the ordinary perturbative realization of the effective action, as defined by the functional integral. 
\subsection{Consistency conditions}
$V_E$ induces only a finite renormalization of the anomaly action. 
In general, the breaking of Weyl invariance generates the anomalous variation
\begin{equation}
 \delta_\sigma \sm=\frac{1}{(4 \pi)^2}\int d^4 x \sqrt{g}\,\sigma\,\left(c_1 R_{\mu\nu\rho\sigma}R^{\mu\nu\rho\sigma} + c_2 R_{\mu\nu}R^{\mu\nu} +c_3 R^2 + c_4\square R\right) 
\end{equation}
which is constrained by the Wess-Zumino consistency condition 
\begin{equation}
\label{WZ}
\left[\delta_{\sigma_1},\delta_{\sigma_2}\right]\sm=0
\end{equation}
to take the form 
\beq
\label{anom1}
\delta_\sigma \sm=\frac{1}{(4\pi)^2}\int d^4 x\sqrt{g}\,\sigma\left( b_1 C^{(4)}_{\mu\nu\rho\sigma}C^{(4)\mu\nu\rho\sigma} + b_2 E_{4} +b_3 \square R\right),
\eeq
in terms of the dimension-4 curvature invariants given above. Obviously, the inclusion of $V_E$ as a counterterm in the 
renormalization procedure is not dictated by the need of canceling any singularity, due to its evanescence at $d=4$.  
Notice that $(C^{(d)})^2$ has a parametric dependence on $d$
\begin{align}
C^{(d)}_{\alpha\beta\gamma\delta} &= R_{\alpha\beta\gamma\delta} -
\frac{4}{d-2}( g_{\alpha(\gamma} \, R_{\delta)\beta} 
- g_{\beta(\gamma} \, R_{\delta)\alpha}) +
\frac{4}{(d-1)(d-2)} \, g_{\alpha(\gamma} \, g_{\delta)\beta}  R\, ,\\
g_{\alpha(\gamma} \, g_{\delta)\beta}&\equiv\frac{1}{2}\left(g_{\alpha\gamma} \, g_{\delta\beta}+g_{\alpha\delta} \, g_{\gamma\beta}\right).
\end{align}
The choice of $(C^{(d)})^2\equiv C^2$ instead of $(C^{(4)})^2$ in the counterterm action takes to variations which are deprived of the total derivative ($\square R$) term in \eqref{anom1}. 
With such a choice of $(C^{(d)})^2$, one derives the relation
\begin{equation}
\frac{\delta}{(d-4)\delta \sigma (x) }\int d^d x \sqrt{-g} (C^{(d)})^2 =\sqrt{-g} (C^{(d)})^2,
\label{oneq}
\end{equation}
which differs from an analogous one
\begin{equation}
\frac{\delta}{(d-4)\delta \sigma (x) }\int d^d x \sqrt{-g} (C^{(4-\epsilon)})^2 =\sqrt{-g}\Bigg( (C^{(4)})^2
+\frac{2}{3}\square R\Bigg) 
\label{twoq1}
\end{equation}
obtained by the replacement of $(C^{(d)})^2\to (C^{(4-\epsilon)})^2$ in the integrand, followed by an expansion of the parametric dependence on $\epsilon$, inducing a finite renormalization of the effective action. Explicitly
\begin{equation}
\label{interm}
(C^{(4-\epsilon)})^2=(C^{(4)})^2 +\epsilon\left( - (R_{\mu\nu})^2 + \frac{5}{18}R^2\right).
\end{equation}
Using the fact that both $C^{(d)}$ and $C^{(4)}$ carry the same Weyl scaling 
\begin{equation}  
C{}_{\lambda\mu\nu\rho}=e^{2 \sigma} \bar{C}_{\lambda\mu\nu\rho}\qquad C^{\lambda\mu\nu\rho}=e^{-6  \sigma} \bar{C}^{ \lambda\mu\nu\rho}
\end{equation}
 one obtains 
\begin{equation} 
\frac{\delta}{\delta \sigma(x)}\int d^d x \sqrt{-g}(C^{4)})^2(x)=\epsilon \sqrt{g}(C^{(4)})^2,
\end{equation}
and using
\begin{equation} 
\frac{\delta}{\delta \sigma(x)}\int d^d x \sqrt{-g}\left( - R_{\mu\nu}^2 + \frac{5}{18}R^2\right)=
-\frac{2}{3}\epsilon \sqrt{g} \Box R
\end{equation}
one finally derives the relation \eqref{twoq1}, with the inclusion of a local (scheme-dependent) term $\Box R$.

\section{ The structure of the effective action from the Conformal Ward identities}
The expansion of $\sm(g)$ can be studied by analising the conformal Ward identities (CWI's) that it has to satisfy. They allow to uncover the structure of the $TT$, $TTT$ - and so on - correlators, quite generally. In particular, for 3-point functions, one can identify the tensorial structures that emerge from the expansion of the $TTT$ either in coordinate or in momentum space, only modulo 3 constants at $d=4$, with similar results holding in other spacetime  dimensions. \\
These results can be derived without the need to resort to a Lagrangian realization, being the CWI's powerful enough for this goal. In this case, free field theory provides a straightforward and complete simplification of the general solution of the CWI's. Notice that this occurs because the scaling dimensions of the stress energy tensor are fixed by the spacetime dimensions $d$. \\
A similar result holds for the conserved currents.  
Correlators with $T$'s and $J$'s have been explicitly discussed in the previous literature, for scalars, fermions and spin-1, both in the free field theory realization and by solving the corresponding CWI's either in coordinate or in momentum space. One could, equivalently, try to determine the structure of such correlation functions by solving directly the CWI's, without any reference to free field theory. However, 
as far as we combine 3 independent sectors with an arbitrary number of fields of those three types 
$(n_S,n_V,n_\psi)$, we reproduce the general solution of the CWI's by matching the constants of such solution $c_i(n_S,n_V,n_\psi)$ \cite{Bzowski:2013sza} with the free field multiplicites \cite{Coriano:2018bsy}. \\
The tensorial structures and the form factors identified by the general solution are identical to those reproduced by the free field theory. At $d=4$ this provides a remarkable simplification of the result.  This matching is pretty useful, since the generalized hypergeometric functions which appear in the general solution can be uniquely expressed in terms of the two master integrals of the perturbative $TTT$, which are the scalar self-energy $B_0$ and the scalar 3-point function $C_0$.\\   
 It is worth to emphasize that the conformal constraints that the correlation functions have to satisfy, can be derived directly from the effective action 
 $\sm(g)$. We are briefly going to review this point since it defines a different formalism, compared to the traditional operatorial one (see for instance \cite{Coriano:2018bbe}) for the derivation of the CWI's. \\
 In general, the variation on the functional $\sm(g)$ generates the relation 
\begin{equation}
\label{anomx}
\frac{\delta \sm}{\delta \sigma(x)}=\sqrt{g} \,g_{\mu\nu}\,\langle T^{\mu\nu}\rangle, 
\qquad \langle T^{\mu\nu}\rangle = \frac{2}{\sqrt{g}}\frac{\delta \sm}{\delta g_{\mu\nu}},
\end{equation} 
and its invariance under Weyl  \begin{equation}
\delta_\sigma g_{\mu\nu}= 2 \sigma g_{\mu\nu} 
\end{equation}
and diffeomorphism transformations
\begin{equation} 
\delta_\epsilon g_{\mu\nu}=-\nabla_{\mu}\epsilon_{\nu}- \nabla_{\nu}\epsilon_{\mu}, 
\end{equation}
gives the conditions
\begin{equation} 
\label{eww}
\delta_\sigma \sm=0 \qquad \delta_\epsilon \sm=0.
\end{equation}
These relations generate trace and conservation constraints on the quantum averages of $T^{\mu\nu}$
 \begin{equation}
\label{comby}
\langle T^\mu_\mu\rangle=0 \qquad \qquad \nabla_\mu\langle T^{\mu\nu}\rangle=0.
\end{equation}
Trace and conformal WI's of the hierarchy of the $nT$'s can be derived from the equations above by functional differentiations of $\sm(g)$ with respect to the metric background. At a second stage, after renormalization, these conditions are modified by the appearance of the conformal anomaly on the right-hand-sides, yielding anomalous CWI's. For example, the anomalous trace WIs can be derived by allowing for an anomaly contribution on the right-hand-side of the $\sigma$ variation in  \eqref{eww}
\begin{equation}
\label{plus}
\delta_\sigma \sm=\int d^4 x\sqrt{g}\,\sigma\,\bar{\mathcal{A}}(x) \qquad \langle T^\mu_\mu\rangle =\bar{\mathcal{A}}(x),
\end{equation}

that clearly violates Weyl invariance. We have denoted with $\sqrt{g}\, \bar{\mathcal{A}}(x)=\sqrt{g}(b E + b' C^2)$ the anomaly. Functional differentiations of this relation generates the hierarchy of trace WIs.\\
The identification of the special CWIs using the formalism of the effective action, can also be addressed in this formulation, by defining currents which are associated to symmetries of $\sm$. An example is the current 
\begin{equation} 
\label{confcur}
\langle J^\mu\rangle =\epsilon_\nu^{(K)} \langle T^{\mu\nu}\rangle, 
\end{equation}
where $\epsilon_\mu^{(K)}(x)$ is a Killing vector field of the metric $g$, which is conserved. The proof follows closely the classical geometric derivation of the conservation of such current. Indeed, we recall that $\epsilon_\mu^{(K)}(x)$ characterizes the isometries of 
$g$
\begin{equation}
 (d s')^2=(d s)^2 \qquad \leftrightarrow  \qquad \nabla_\mu\epsilon_{\nu}^{(K)} + \nabla_\nu\epsilon_{\mu}^{(K)}=0.
\end{equation}
Then, the requirement of diffeomorphism invariance of the effective action $\sm(g)$, summarised by the second condition in \eqref{comby},
implies the conservation equation
\begin{equation}
\label{iso}
\nabla\cdot \langle J\rangle =0. 
\end{equation}
Such equation can be re-obtained  by requiring the invariance of $\sm$ under a specific variation respect to Killing vectors (KVs)  $\epsilon_\mu^{(K)}$ of the form
\begin{equation}
\delta_{KV}\sm=0.  
\end{equation}
If we require that the metric background allows conformal Killing vectors (CKVs) and the effective action is invariant under such transformations, then we have the possibility of taking into account the anomaly contribution to the equations.  \\
We recall that the CKVs are solution of the equation
\begin{equation}
(ds')^2 = e^{2 \sigma(x)}(ds)^2 \qquad \leftrightarrow\qquad  \nabla_\mu \epsilon_\nu^{(K)} + \nabla_\nu \epsilon_\mu^{(K)} = 2 \sigma \delta_{\mu\nu}\qquad \sigma=\frac{1}{4}\nabla\cdot\epsilon^{(K)}.
\end{equation}
Notice that if we introduce a conformal current $J_c$, defined as in \eqref{confcur} but now using the CKVs of the background metric, if conditions \eqref{comby} are respected by $\sm$, then $J_c$ is conserved as in the isometric case \eqref{iso}. \\
Things are slightly different if we allow for a Weyl-variant term in $\sm$ as in \eqref{plus}, which takes place in $d=4$, after renormalization. \\
In this case the anomaly induces a nonzero trace, and modifies the semiclassical condition \eqref{iso} into the new form
\begin{equation}
\nabla\cdot \langle J_c\rangle =\frac{1}{4}\nabla \cdot \epsilon^{(K)} \langle T^\mu_\mu\rangle  +\epsilon^{(K)}_\nu \nabla_{\mu} \langle T^{\mu\nu}\rangle.
\end{equation}
This relation can be imposed as a conservation equation on correlators of various forms, in order to extract the CWI's. For instance, these correlators can be chosen of the form $J_c T$ in $d$ dimensions. As far as we stay in $d$ dimensions, there are no anomalies to modify the constraints, being them induced only at $d=4$. Notice that $\sigma(x)$ in \eqref{plus} is, at the beginning, a generic scalar function, that can be Taylor expanded around a given point $x^\mu$, and is characterized, in principle, by an infinite and arbitrary number of constants. Obviously, 
the number of such constants gets drastically reduced if we require that the spacetime manifold allows a tangent space (a free-falling coordinate frame) at each of its points, endowed with a flat conformal symmetry. \\
Indeed, in flat space, the conformal Killing equation is solved by some  CKVs $\epsilon^\mu$, which are at most quadratic in $x$, and are expressed in terms of the 15 parameters $(a^\mu,\omega^{\mu\nu}, \lambda_s, b^\mu)$ of the conformal group, indicated as $K^\mu(x)$ 
\begin{equation}
\label{Kil}
\epsilon^{\mu}(x)\,\big|_{flat}\to K^\mu(x)= a^\alpha + \omega^{\mu\nu} x^\nu +\lambda_s x^\mu + b^\mu x^2 -2 x^\mu b\cdot x.
\end{equation}
Using such CKVs, the derivation of the CWIs, following the approach of \cite{Coriano:2017mux}, can be performed directly in $d=4$. We are going to illustrate this derivation for 4-point functions in the next section.

\subsection{The anomalous CWI's using conformal Killing vectors}
\label{heres}
The expressions of the anomalous conformal WIs can be derived in an alternative way following the formulation of  \cite{Coriano:2017mux}, that here we are going to discuss for the 4-graviton vertex (TTTT). As just mentioned, the derivation of such identities relies uniquely on the effective action. We illustrate it first in the TT case, and then move to the 4T.\\
We start from the conservation of the conformal current as derived in \eqref{iso}
\begin{equation}
\int d^d x \sqrt{g}\, \nabla^\alpha \left( \epsilon_\alpha \frac{2}{\sqrt{g}}\frac{\delta \sm}{\delta g_{\mu\alpha}}\right)=\int d^4 x \sqrt{g}\, \nabla_\mu\braket{\epsilon_\alpha T^{\alpha \nu}} =0.  
\end{equation}
In the TT case the derivation of the special CWIs is simplified, since there is no trace anomaly if the counterterm action is defined as in \eqref{counter}. We rely on the fact that the conservation of the conformal current $J^\mu_{(K)}$ implies the conservation equation 
\begin{align}
0=\int\,d^dx\,\sqrt{-g}\, \,\nabla_\mu\,\braket{J^\mu_{(K)}(x)\,T^{\mu_1\nu_1}(x_1)}.
\end{align}
By making explicit the expression $J^\mu(x)=K_\nu(x)\,T^{\mu\nu}(x)$, with $\epsilon \to K$ in the flat limit,  the previous relation takes the form
 \begin{align}
0=\int\,d^dx\,\bigg(\partial_\mu K_\nu\,\braket{T^{\mu\nu}(x)\,T^{\mu_1\nu_1}(x_1)}+ K_\nu\,\partial_\mu\,\braket{T^{\mu\nu}(x)\,T^{\mu_1\nu_1}(x_1)}\bigg).\label{cons}
 \end{align}
We recall that $K_\nu$ satisfies the conformal Killing equation in flat space
\begin{align}
\label{flatc}
\partial_\mu K_\nu+\partial_\nu K_\mu=\frac{2}{d}\delta_{\mu\nu}\,\left(\partial\cdot K\right),
\end{align}
and by using this equation \eqref{cons} can be re-written in the form
\begin{align}
	0=\int\,d^dx\,\bigg(K_\nu\partial_\mu\,\braket{T^{\mu\nu}(x)\,T^{\mu_1\nu_1}(x_1)}+\frac{1}{d}\big(\partial\cdot K\big)\,\braket{T(x)\,T^{\mu_1\nu_1}(x_1)}\bigg).\label{newcons}
\end{align}
We can use in  this previous expression the conservation and trace Ward identities for the two-point function $\braket{TT}$, that in the flat spacetime limit are explicitly given by
\begin{align}
\partial_\mu\braket{T^{\mu\nu}(x)T^{\mu_1\nu_1}(x_1)}&=\bigg(\delta^{(\mu_1}_\mu\delta^{\nu_1)}_\lambda\partial^\nu\delta(x-x_1)-2\delta^{\nu(\mu_1}\delta^{\nu_1)}_\mu\partial_\lambda\delta(x-x_1)\bigg)\braket{T^{\lambda\mu}(x)},\label{consTT}\\
\delta_{\mu\nu}\braket{T^{\mu\nu}(x)T^{\mu_1\nu_1}(x_1)}&\equiv\braket{T(x)T^{\mu_1\nu_1}(x_1)}=-2\delta(x-x_1)\braket{T^{\mu_1\nu_1}(x)}\label{traceTT}
\end{align}
and the explicit expression of the Killing vector $K^{(C)}_\nu$ for the special conformal transformations 
\begin{equation}
\begin{split}
K^{(C)\,\kappa}_\mu&=2x^\kappa\,x_\mu-x^2\delta^\kappa_\mu\\
\partial\cdot  K^{(C)\,\kappa}&=2d\,x^\kappa
\end{split}\label{spc}
\end{equation}
where $\kappa=1,\dots,d$. By using \eqref{spc} in the integral \eqref{newcons}, we can rewrite that expression as
\begin{align}
	0=\int\,d^dx\,\bigg[\big(2x^\kappa\,x_\nu-x^2\delta^\kappa_\nu\big)\partial_\mu\,\braket{T^{\mu\nu}(x)\,T^{\mu_1\nu_1}(x_1)}+2\,x^\kappa\,\braket{T(x)\,T^{\mu_1\nu_1}(x_1)}\bigg],
\end{align}
We need just a final integrating by parts to finally obtain the special CWI relation
\begin{align}
&\left(2d\,x_1^\kappa+2x_1^\kappa\,x^{\mu}_1\frac{\partial}{\partial x_1^\mu}+x_1^2\frac{\partial}{\partial x_{1\kappa}}\right)\braket{T^{\mu_1\nu_1}(x_1)}\notag\\
&\quad+2\bigg(x_{1\lambda}\,\delta^{\mu_1\kappa}-x_1^{\mu_1}\delta^\kappa_\lambda\bigg)\braket{T^{\lambda\nu_1}(x_1)}+2\bigg(x_{1\lambda}\,\delta^{\nu_1\kappa}-x_1^{\nu_1}\delta^\kappa_\lambda\bigg)\braket{T^{\mu_1\lambda}(x_1)}=0
\end{align}
 for the 1-point function $\braket{T^{\mu_1\nu_1}(x_1)}$. 
\subsection{ 4-point functions}\label{heres2}
The derivation above can be extended to n-point functions, starting from the identity 
\begin{equation} 
\int d^d x \sqrt{g} \nabla_\alpha(x) \langle J^\alpha_c(x)T^{\mu_1\nu_1}(x_1)\ldots T^{\mu_n\nu_n}(x_n)
\label{div}
\rangle=0.
\end{equation}
We have used the conservation of the conformal current in d dimensions under variations of the metric, induced by the conformal Killing vectors.\\
In the absence of an anomaly, the conservation of the current $J^\mu_c$ follows from the conservation of the stress energy tensor, plus the zero trace condition.
As in the example illustrated above, we consider \eqref{div} in the flat limit 
\begin{align}
\int dx^d \,\partial_\nu\bigg[K_\mu(x)\braket{T^{\mu\nu}(x)T^{\mu_1\nu_1}(x_1)\dots T^{\mu_4\nu_4}(x_4)}\bigg]=0,\label{Killing1}
\end{align}
where we are assuming that the surface terms vanish, due to the fast fall-off behaviour of the correlation function at infinity. Expanding \eqref{Killing1} we obtain an expression similar to \eqref{newcons}

\begin{align}
0=\int d^dx\left\{K_{\mu}(x)\partial_\nu\braket{T^{\mu\nu}(x)T^{\mu_1\nu_1}(x_1)\dots T^{\mu_4\nu_4}(x_4)}+\frac{1}{d}\big(\partial\cdot K\big)\delta_{\mu\nu}\braket{T^{\mu\nu}(x)T^{\mu_1\nu_1}(x_1)\dots T^{\mu_4\nu_4}(x_4)}
\right\}.\label{killing01}
\end{align}
Starting  from this expression, the dilatation CWI is obtained by the choice of the CKV characterising the dilatations  
\begin{equation}
K^{(D)}_\mu(x)=x_\mu,\qquad\partial\cdot K^{(D)}=d
\end{equation}
and \eqref{killing01} becomes
\begin{align}
0=\int d^d x\bigg\{x_\mu\,\partial_\nu\braket{T^{\mu\nu}(x)T^{\mu_1\nu_1}(x_1)\dots T^{\mu_4\nu_4}(x_4)}+\delta_{\mu\nu}\braket{T^{\mu\nu}(x)T^{\mu_1\nu_1}(x_1)\dots T^{\mu_4\nu_4}(x_4)}\label{anomDil}
\bigg\}.
\end{align}
At this stage, we use the conservation and trace Ward identities in $d=4$ for the $4$-point function written as
\begin{align}
&\partial_\nu\braket{T^{\mu\nu}(x)T^{\mu_1\nu_1}(x_1)\dots T^{\mu_4\nu_4}(x_4)}=\notag\\
=&-8\bigg\{\left[\Gamma^{\mu}_{\nu\lambda}(x)\right]^{\mu_1\nu_1\mu_2\nu_2\mu_3\nu_3}(x_1,x_2,x_3)\braket{T^{\lambda\nu}(x)T^{\mu_4\nu_4}(x_4)}+(14)+(24)+(34)\bigg\}\notag\\
&-4\bigg\{\left[\Gamma^{\mu}_{\nu\lambda}(x)\right]^{\mu_1\nu_1\mu_2\nu_2}(x_1,x_2)\braket{T^{\lambda\nu}(x)T^{\mu_3\nu_3}(x_3)T^{\mu_4\nu_4}(x_4)}+(13)+(23)+(14)+(24)+(34)\bigg\}\notag\\
&-2\bigg\{\left[\Gamma^{\mu}_{\nu\lambda}(x)\right]^{\mu_1\nu_1}(x_1)\braket{T^{\lambda\nu}(x)T^{\mu_2\nu_2}(x_2)T^{\mu_3\nu_3}(x_3)T^{\mu_4\nu_4}(x_4)}+(12)+(13)+(14)\bigg\}\label{5ptcons}
\end{align}
and
\begin{align}
&\delta_{\mu\nu}\braket{T^{\mu\nu}(x)T^{\mu_1\nu_1}(x_1)\dots T^{\mu_4\nu_4}(x_4)}\notag\\
&\hspace{1cm}=-2\bigg\{\delta_{xx_1}\braket{T^{\mu_1\nu_1}(x)T^{\mu_2\nu_2}(x_2)\dots T^{\mu_4\nu_4}(x_4)}+(12)+(13)+(14)\bigg\}\notag\\
&\hspace{2.5cm}+2^4\big[\mathcal{A}(x)\big]^{\mu_1\nu_1\dots\mu_4\nu_4}(x_1,\dots,x_4).\label{5pttrace}
\end{align}
to finally derive the dilatation WI from \eqref{anomDil} in the form
\begin{align}
\left(4d+\sum_{j=1}^4\,x_j^\alpha\frac{\partial}{\partial x_j^\alpha}\right)\braket{T^{\mu_1\nu_1}(x_1)\dots T^{\mu_4\nu_4}(x_4)}=2^4\int dx \big[\mathcal{A}(x)\big]^{\mu_1\nu_1\dots\mu_4\nu_4}(x_1,\dots,x_4),\label{DilAnom}
\end{align}
where $d=4$. It is worth mentioning that \eqref{DilAnom} is valid in any even spacetime dimension if we take into account the particular structure of the trace anomaly in that particular dimension.

The special CWIs correspond to the $d$ special conformal Killing vectors in flat space given in \eqref{spc}, as in the $TT$ case. Also in this case we derive the identity 
\begin{align}
0=&\int d^d x\bigg\{\big(2x^\kappa\,x_\mu-x^2\delta^\kappa_\mu\big)\,\partial_\nu\braket{T^{\mu\nu}(x)T^{\mu_1\nu_1}(x_1)\dots T^{\mu_4\nu_4}(x_4)}\notag\\
&\hspace{4cm}+2x^\kappa\delta_{\mu\nu}\braket{T^{\mu\nu}(x)T^{\mu_1\nu_1}(x_1)\dots T^{\mu_4\nu_4}(x_4)}\label{anomSpecWI}
\bigg\}.
\end{align}
By using the relations \eqref{5ptcons} and \eqref{5pttrace} and performing the integration over $x$ explicitly in the equation above,  the anomalous special CWIs for the $4$-point function take the form
\begin{align}
&\sum_{j=1}^4\left[2x_j^\kappa\left(d+x_j^\alpha\frac{\partial}{\partial x_j^\alpha}\right)-x_j^2\,\delta^{\kappa\alpha}\frac{\partial}{\partial x_j^\alpha}\right]\braket{T^{\mu_1\nu_1}(x_1)\dots T^{\mu_4\nu_4}(x_4)}\notag\\
&+2\sum_{j=1}^4\left(\delta^{\kappa\mu_j}x_{j\,\alpha}-\delta^\kappa_\alpha x_j^{\mu_j}\right)\braket{T^{\mu_1\nu_1}(x_1)\dots T^{\nu_j\alpha}(x_j)\dots T^{\mu_4\nu_4}(x_4)}\notag\\
&+2\sum_{j=1}^4\left(\delta^{\kappa\nu_j}x_{j\,\alpha}-\delta^\kappa_\alpha x_j^{\nu_j}\right)\braket{T^{\mu_1\nu_1}(x_1)\dots T^{\mu_j\alpha}(x_j)\dots T^{\mu_4\nu_4}(x_4)}=\notag\\
&=2^5\,\int dx\,x^\kappa\big[\mathcal{A}(x)\big]^{\mu_1\nu_1\dots\mu_4\nu_4}(x_1,\dots,x_4),
\end{align}
where the presence of the anomaly term comes from the inclusion of the trace  WI, exactly as in the TT case. 
\section{Conservation Ward identities}
Similar constraints are enforced from the invariance of $\sm{(g)}$ under diffeomorphisms 

\begin{equation}
\label{vat}
^{(g)}\nabla_\mu\braket{T^{\mu\nu}(x)}_g=0, \qquad i.e. \qquad \delta_\epsilon \sm(g)=0
\end{equation}

Here $^{(g)}\nabla_\mu$ denotes the covariant derivative in the general background metric $g_{\mu\nu}(x)$. It can be expressed in the form 
\begin{equation}
\partial_\nu\left(\frac{\delta \sm(g)}{\delta g_{\mu\nu}(x)}\right)+\sm^\mu_{\nu\lambda}\left(\frac{\delta \sm(g)}{\delta g_{\lambda\nu}(x)}\right)=0\label{conserv},
\end{equation}
where $\Gamma^\mu_{\lambda\nu}$ is the Christoffel connection for the general background metric $g_{\mu\nu}(x)$.\\

In order to derive the conservation WIs for higher point correlation functions, one has to consider additional variations with respect to the metric of \eqref{vat} and then move to flat space, obtaining 
\begin{align}
&\partial_{\nu_1}\braket{T^{\mu_1\nu_1}(x_1)T^{\mu_2\nu_2}(x_2)T^{\mu_3\nu_3}(x_3)T^{\mu_4\nu_4}(x_4)}=\notag\\
=&-\left[2\left(\frac{\delta\Gamma^{\mu_1}_{\lambda\nu_1}(x_1)}{\delta  g_{\mu_2\nu_2}(x_2)}\right)_{g=\delta}\braket{T^{\lambda\nu_1}(x_1)T^{\mu_3\nu_3}(x_3)T^{\mu_4\nu_4}(x_4)}+(23)+(24)\right]\notag\\
& -\left[4\left(\frac{\delta^2\Gamma^{\mu_1}_{\lambda\nu_1}(x_1)}{\delta  g_{\mu_2\nu_2}(x_2)\delta  g_{\mu_3\nu_3}(x_3)}\right)_{g=\delta}\braket{T^{\lambda\nu_1}(x_1)T^{\mu_4\nu_4}(x_4)}+(24)+(34)\right],\label{transverseX}
\end{align}
where 
\begin{align}
\left(\frac{\delta\Gamma^{\mu_1}_{\lambda\nu_1}(x_1)}{\delta  g_{\mu_i\nu_i}(x_i)}\right)_{g=\delta}&=\frac{1}{2}\left(\delta^{\mu_1(\mu_i}\delta^{\nu_i)}_{\nu_1}\,\partial_\lambda\delta_{x_1x_i}+\delta^{\mu_1(\mu_i}\delta^{\nu_i)}_{\lambda}\,\partial_{\nu_1}\delta_{x_1x_i}-\delta^{(\mu_i}_\lambda\delta^{\nu_i)}_{\nu_1}\,\partial^{\mu_1}\delta_{x_1x_i}\right)\\[1.5ex]
\left(\frac{\delta^2\Gamma^{\mu_1}_{\lambda\nu_1}(x_1)}{\delta  g_{\mu_i\nu_i}(x_i)\delta  g_{\mu_j\nu_j}(x_j)}\right)_{g=\delta}&=\notag\\
&\hspace{-3cm}=-\frac{\delta_{x_1x_i}}{2}\delta^{\mu_1(\mu_i}\delta^{\nu_i)\epsilon}\left(\delta^{(\mu_j}_\epsilon\delta^{\nu_j)}_{\nu_1}\,\partial_\lambda\delta_{x_1x_j}+\delta^{(\mu_j}_\epsilon\delta^{\nu_j)}_{\lambda}\,\partial_{\nu_1}\delta_{x_1x_j}-\delta^{(\mu_j}_\lambda\delta^{\nu_j)}_{\nu_1}\,\partial_\epsilon\delta_{x_1x_j}\right)+(ij),
\end{align}
are the first and second functional derivatives of the connection, in the flat limit.
We have explicitly indicated the symmetrization with respect to the relevant indices using the permutation $(ij)\equiv (i\leftrightarrow j)$.  We have introduced a simplified notation for the Dirac delta $\delta_{x_ix_j}\equiv \delta(x_i-x_j)$. All the derivative (e.g. $\partial_\lambda$) are taken with respect to the coordinate $x_1$  $(e.g. \partial/\partial x_1^\lambda)$. \\
We adopt the convention
\bea
\braket{T^{\mu_1\nu_1}(p_1)T^{\mu_2\nu_2}(p_2)T^{\mu_3\nu_3}(p_3)T^{\mu_4\nu_4}( \overline{p}_4)}
&=&\int d^4 x_1 d^4 x_2 d^4 x_3 e^{-i( p_1\cdot x_1 +p_2\cdot x_2 + p_3\cdot x_3)}\nonumber \\
&& \times \braket{T^{\mu_1\nu_1}(x_1)T^{\mu_2\nu_2}(x_2)T^{\mu_3\nu_3}(x_3)T^{\mu_4\nu_4}(0)}
\eea
to transform to momentum space and use the translational invariance of the correlator in flat space, which allows to use momentum conservation to express one of the momenta (in our convention this is $p_4$) as combination of the remaining ones $\bar{p}_4=-p_1-p_2-p_3$. Details on the elimination of one of the momenta in the derivation of the CWIs can be found in \cite{Coriano:2018bbe}. \\
The conservation Ward Identity \eqref{transverseX} in flat spacetime may be Fourier transformed, giving the CWIs in momentum space
\begin{align}
&p_{1\nu_1}\braket{T^{\mu_1\nu_1}(p_1)T^{\mu_2\nu_2}(p_2)T^{\mu_3\nu_3}(p_3)T^{\mu_4\nu_4}(\bar p_4)}=\notag\\
&=\Big[4\, \mathcal{B}^{\mu_1\hspace{0.4cm}\mu_2\nu_2\mu_3\nu_3}_{\hspace{0.3cm}\lambda\nu_1}(p_2,p_3)\braket{T^{\lambda\nu_1}(p_1+p_2+p_3)T^{\mu_4\nu_4}(\bar p_4)}+(34)+(2 4)\Big]\notag\\
&\hspace{0.5cm}+\Big[2 \, \mathcal{C}^{\mu_1\hspace{0.4cm}\mu_2\nu_2}_{\hspace{0.3cm}\lambda\nu_1}(p_2)\braket{T^{\lambda\nu_1}(p_1+p_2)T^{\mu_3\nu_3}(p_3)T^{\mu_4\nu_4}(\bar p_4)}+(2 3)+(2 4)\Big],\label{transverseP}
\end{align}
where we have defined
\begin{align}
\label{BB}
\mathcal{B}^{\mu_1\hspace{0.4cm}\mu_2\nu_2\mu_3\nu_3}_{\hspace{0.3cm}\lambda\nu_1}(p_2,p_3)&\equiv -\frac{1}{2}\delta^{\mu_1(\mu_2}{\delta^{\nu_2)\epsilon}}\left(\delta_\epsilon^{(\mu_3}\delta^{\nu_3)}_{\nu_1}\,p_{3\,\lambda}+\delta_\epsilon^{(\mu_3}\delta^{\nu_3)}_{\lambda}\,p_{3\,\nu_1}-\delta_\lambda^{(\mu_3}\delta^{\nu_3)}_{\nu_1}\,p_{3\,\epsilon}\right)+(23)\\[2ex]
\label{CC}
\mathcal{C}^{\mu_1\hspace{0.4cm}\mu_2\nu_2}_{\hspace{0.3cm}\lambda\nu_1}(p_2)&\equiv \frac{1}{2}\left(\delta^{\mu_1(\mu_2}\delta^{\nu_2)}_{\nu_1}\,p_{2\,\lambda}+\delta^{\mu_1(\mu_2}\delta^{\nu_2)}_{\lambda}p_{2\,\nu_1}-\delta^{(\mu_2}_{\lambda}\delta^{\nu_2)}_{\nu_1}p_2^{\mu_1}\right),
\end{align}
related to the second and first functional derivatives of the Christoffel connection, respectively.

\section{Renormalization in momentum space around Minkowski}
To investigate the renormalized correlator and the anomaly-induced effective action, one may just focus on the counterterm action.  
For a generic $nT$ correlator, the only counterterm needed for its renormalization, is obtained by the inclusion of a classical gravitational vertex generated by the differentiation of the action 
\begin{align}
\label{counter}
\sm_{ct}(g)&=-\frac{1}{\epsilon}(b' V_E + b\, V_{C^2})
\end{align}

 $n$ times.\\
The renormalized effective action $\sm_R$ is then defined by the sum of the two terms
\begin{equation}
\sm_R(g)=\sm(g)+ \sm_{ct}(g)
\end{equation}
\begin{align}
\label{cct1}
\sm_{ct}=\raisebox{-0.8ex}{\includegraphics[width=0.15\linewidth]{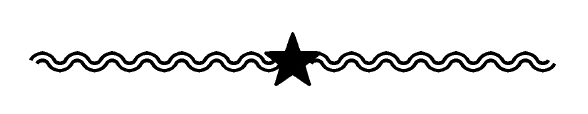}}+\raisebox{-5ex}{\includegraphics[width=0.15\linewidth]{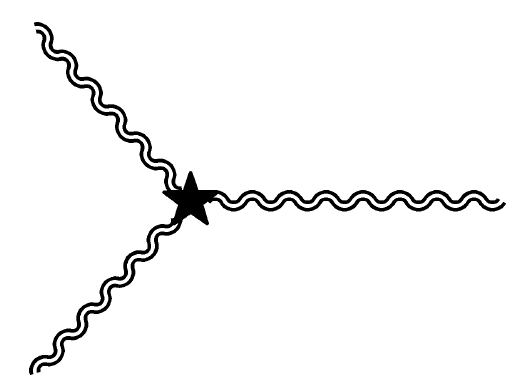}}+\raisebox{-6ex}{\includegraphics[width=0.15\linewidth]{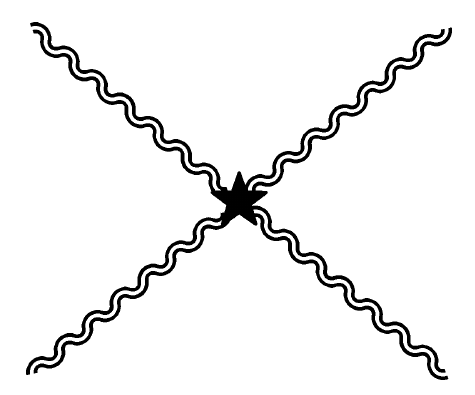}}+...
\end{align}
with $\sm_{ct}$ shown in \eqref{cct1}. Both terms of $\sm_R(g)$ are expanded in the metric fluctuations as in \eqref{cct1}. 
The correlation functions obrtained from the renormalized action can be expressed as the sum of a finite $(f)$  correlator and of an anomaly term ($anomaly$) in the form
\begin{align}
\label{cct}
 \braket{T^{\mu_1\nu_1}T^{\mu_2\nu_2}\ldots T^{\mu_n\nu_n}}_{Ren}=
&\bigg[\braket{T^{\mu_1\nu_1}T^{\mu_2\nu_2}\ldots T^{\mu_n\nu_n}}_{bare}+\braket{T^{\mu_1\nu_1}T^{\mu_2\nu_2}\ldots T^{\mu_n\nu_n}}_{count}\bigg]_{d\to4}=\notag\\
&=\braket{T^{\mu_1\nu_1}T^{\mu_2\nu_2}\ldots T^{\mu_n\nu_n}}^{(d=4)}_{f}+\braket{T^{\mu_1\nu_1}T^{\mu_2\nu_2}\ldots T^{\mu_n\nu_n}}^{(d=4)}_{anomaly}
\end{align} 
and satisfy anomalous CWI's.
To characterize the anomaly contribution to each correlation function, we start from the 1-point function.\\
 In a generic background $g$, the 1-point function is decomposed as 
\begin{equation}
\label{decomp}
\langle T^{\mu\nu}\rangle_{Ren}=\frac{2}{\sqrt{g}}\frac{\delta \sm_{Ren}}{\delta g_{\mu\nu}} =\langle T^{\mu\nu} \rangle_A  + \langle \overline{T}^{\mu\nu}\rangle_f,
\end{equation}
with
\begin{equation}
g^{\mu\nu}\frac{\delta \sm}{\delta g^{\mu\nu}} = g^{\mu\nu}\frac{\delta \sm_A}{\delta g^{\mu\nu}}\equiv \frac{\sqrt{g}}{2} g_{\mu\nu} \langle T^{\mu\nu} \rangle_A \end{equation}
being the trace anomaly equation, and $\langle \overline{T}^{\mu\nu}\rangle_f$ is the Weyl-invariant (traceless) term.
\\
Following the discussion in \eqref{anom1}, these scaling violations may be written for the 1-point function in the form
\begin{equation}
\label{anomeq}
\braket{T^{\mu}_{\ \ \mu}(x)}=\mathcal{A}(x)
\end{equation} 
 with the finite terms on the right hand side of this equation denote the anomaly contribution
\begin{equation}
\label{AF}
\mathcal{A}(x)=\sqrt{-g(x)}\bigg[b\,C^2(x)+b'E(x)\bigg]
\end{equation}
The extraction of the anomalous CWI's around flat space requires several differentiation of this functional relation, 
evaluated in the flat limit. This procedure generates, for the anomaly contribution, expressions which are  polynomial in the momenta.  In general, one also finds additional dimension-4 local invariants $\mathcal{L}_i$, if there are couplings to other background fields, as for instance in the QED and QCD cases, with coefficients related to the $\beta$ functions of the corresponding gauge couplings. \\
We recall that the counterterm vertex for the nT correlator, in DR, in momentum space takes the form
\begin{equation}
\braket{T^{\mu_1\nu_1}(p_1)\dots T^{\mu_n\nu_n}(\bar{p}_n)}_{count}=-\frac{\mu^{-\varepsilon}}{\varepsilon}\bigg(b\,V_{C^2}^{\mu_1\nu_1\dots\mu_n\nu_n}(p_1,\dots,\bar{p}_n)+b'\,V_{E}^{\mu_1\nu_1\dots\mu_n\nu_n}(p_1,\dots,\bar{p}_n)\bigg)
\label{nTcount},
\end{equation}
 where 
 
\begin{align}
V_{C^2}^{\mu_1\nu_1\dots\mu_n\nu_n}(p_1,\dots,\bar{p}_n)
&\equiv 2^n\big[\sqrt{-g}\,C^2\big]^{\mu_1\nu_1\dots\mu_n\nu_n}(p_1,\dots,\bar{p}_n)\notag\\
&=2^n\int\,d^dx_1\,\dots\,d^dx_n\,d^dx\,\bigg(\sdfrac{\delta^n(\sqrt{-g}C^2)(x)}{\delta g_{\mu_1\nu_1}(x_1)\dots\delta g_{\mu_n\nu_n}(x_n)}\bigg)_{g=\delta}\,e^{-i(p_1\,x_1+\dots+p_n x_n)}\notag\\[2ex]
\end{align}
and
\begin{align}
V_{E}^{\m_1\n_1\dots\mu_n\nu_n}(p_1,\dots,\bar{p}_n)&\equiv 2^n\big[\sqrt{-g}\,E\big]^{\m_1\n_1\dots\m_n\n_n}(p_1,\dots,\bar{p}_n)\notag\\
\end{align}
are the expressions of the two contributions present in \eqref{nTcount} in momentum space. 
One can also verify the following trace relations 
\begin{align}
&\delta_{\mu_1\nu_1}\,V_{C^2}^{\m_1\n_1\dots\mu_n\nu_n}(p_1,\dots,p_n)=2^{n-1}(d-4)\,\left[\sqrt{-g}C^2\right]^{\mu_2\nu_2\dots\mu_n\nu_n}(p_2,\dots,p_n)\notag\\
&\hspace{2cm}-2\bigg[V_{C^2}^{\m_2\n_2\dots\mu_n\nu_n}(p_1+p_2,p_3,\dots,p_n)+\dots+V_{C^2}^{\m_2\n_2\dots\mu_n\nu_n}(p_2,p_3,\dots,p_1+p_n)\bigg]\label{expabove}\\[2ex]
&\delta_{\mu_1\nu_1}\,V_{E}^{\m_1\n_1\dots\mu_n\nu_n}(p_1,\dots,p_n)=2^{n-1}(d-4)\,\left[\sqrt{-g}E\right]^{\mu_2\nu_2\dots\mu_n\nu_n}(p_2,\dots,p_n)\notag\\
&\hspace{2cm}-2\bigg[V_{E}^{\m_2\n_2\dots\mu_n\nu_n}(p_1+p_2,p_3,\dots,p_n)+\dots+V_{E}^{\m_2\n_2\dots\mu_n\nu_n}(p_2,p_3,\dots,p_1+p_n)\bigg]\label{mom}
\end{align}
that hold in general $d$ dimensions. \\
Obviously, the effective action that results from the renormalization can be clearly separated in terms of two contributions, as evident from \eqref{decomp}, 
\begin{equation} 
\sm_R(g)=\sm_{A}(g) +{\sm}_f(g),
\end{equation}
corresponding to an anomaly part $\sm_{A}[g]$ and to a finite, Weyl-invariant term, which can be expanded in terms of fluctuations over a background $\bar{g}$ as for the entire effective action $\sm(g)$. This functional collects finite correlators of the form \eqref{cct} in $d=4$, where we find the expansion for $\sm_{A}$, the anomaly part.\\ 
Notice that the anomaly effective action $\sm_A$ obtained from this analysis in momentum space is a rational function of the external momenta, characterised by well-defined tensor structures. We refer to the analysis of the  TJJ and 3T correlators \cite{Giannotti:2008cv,Armillis:2009pq,Armillis:2010qk} for more details. \\
\subsection{Conservation WI's for the counterterms} 
To illustrate the conservation WI in detail, we turn to the expression of the counterterm action \eqref{counter}, which generates counterterm vertices of the form
\begin{align}
&\braket{T^{\mu_1\nu_1}(p_1)T^{\mu_2\nu_2}(p_2)T^{\mu_3\nu_3}(p_3)T^{\mu_4\nu_4}(\bar{p}_4)}_{count}=\notag\\
&\qquad=-\frac{\mu^{-\varepsilon}}{\varepsilon}\bigg(b\,V_{C^2}^{\mu_1\nu_1\mu_2\nu_2\mu_3\nu_3\mu_4\nu_4}(p_1,p_2,p_3,\bar{p}_4)+b'\,V_{E}^{\mu_1\nu_1\mu_2\nu_2\mu_3\nu_3\mu_4\nu_4}(p_1,p_2,p_3,\bar{p}_4)\bigg),\label{TTTTcount}
\end{align}
where on the rhs of the expression above we have introduced the counterterm vertices (with $P= p_1 +\ldots p_4$)
\begin{align}
&V_{C^2}^{\m_1\n_1\m_2\n_2\m_3\n_3\mu_4\nu_4}(p_1,p_2,p_3,\bar{p}_4)
\,\delta^4(P)\equiv 16\, \delta^4(P)\,\big[\sqrt{-g}\,C^2\big]^{\m_1\n_1\m_2\n_2\m_3\n_3\mu_4\nu_4}(p_1,p_2,p_3,\bar{p}_4)\notag\\
&V_{E}^{\m_1\n_1\m_2\n_2\m_3\n_3\mu_4\nu_4}(p_1,p_2,p_3,\bar{p}_4)\delta^4(P)\equiv 16\,\,\delta^4(P)\big[\sqrt{-g}\,E\big]^{\m_1\n_1\m_2\n_2\m_3\n_3\m_4\n_4}(p_1,p_2,p_3,\bar{p}_4)\notag\\
\end{align}
evaluated in the flat spacetime limit.
These vertices share some properties when contracted with flat metric tensors and the external momenta. In particular, from \eqref{expabove} and \eqref{mom}, when $n=4$ and in $d$ dimensions we have
\begin{align}
\delta_{\mu_1\nu_1}\,V_{C^2}^{\m_1\n_1\m_2\n_2\m_3\n_3\mu_4\nu_4}(p_1,p_2,p_3,\bar{p}_4)&=8(d-4)\,\left[\sqrt{-g}C^2\right]^{\mu_2\nu_2\mu_3\nu_3\mu_4\nu_4}(p_2,p_3,\bar{p}_4)\notag\\
&\hspace{-5.5cm}-2V_{C^2}^{\m_2\n_2\m_3\n_3\mu_4\nu_4}(p_1+p_2,p_3,\bar{p}_4)-2V_{C^2}^{\m_2\n_2\m_3\n_3\mu_4\nu_4}(p_2,p_1+p_3,\bar{p}_4)\notag\\
&\hspace{-5cm}-2V_{C^2}^{\m_2\n_2\mu_3\nu_3\m_4\n_4}(p_2,p_3,p_1+\bar{p}_4),\\[2ex]
\delta_{\mu_1\nu_1}\,V_{E}^{\m_1\n_1\m_2\n_2\m_3\n_3\mu_4\nu_4}(p_1,p_2,p_3,\bar{p}_4)&=8(d-4)\,\left[\sqrt{-g}E\right]^{\mu_2\nu_2\mu_3\nu_3\mu_4\nu_4}(p_2,p_3,\bar{p}_4)\notag\\
&\hspace{-5.5cm}-2V_{E}^{\m_2\n_2\m_3\n_3\mu_4\nu_4}(p_1+p_2,p_3,\bar{p}_4)-2V_{E}^{\m_2\n_2\m_3\n_3\mu_4\nu_4}(p_2,p_1+p_3,\bar{p}_4)\notag\\
&\hspace{-5cm}-2V_{E}^{\m_2\n_2\mu_3\nu_3\m_4\n_4}(p_2,p_3,p_1+\bar{p}_4),
\end{align}
which play a key role in the renormalization procedure.
Furthermore, the contraction of these vertices with the external momenta generates conservation WIs in $d$ dimensions, similar to \eqref{transverseP}, 
\begin{align}
&p_{1\,\nu_1}\,V_{C^2}^{\m_1\n_1\m_2\n_2\m_3\n_3\mu_4\nu_4}(p_1,p_2,p_3,\bar{p}_4)=\notag\\
&=\Big[4\, \mathcal{B}^{\mu_1\hspace{0.4cm}\mu_2\nu_2\mu_3\nu_3}_{\hspace{0.3cm}\lambda\nu_1}(p_2,p_3)V_{C^2}^{\lambda\n_1\mu_4\nu_4}(p_1+p_2+p_3,\bar{p}_4)+(34)+ (24)\Big]\notag\\
&\hspace{0.5cm}+\Big[2 \, \mathcal{C}^{\mu_1\hspace{0.4cm}\mu_2\nu_2}_{\hspace{0.3cm}\lambda\nu_1}(p_2)V_{C^2}^{\lambda\n_1\m_3\n_3\mu_4\nu_4}(p_1+p_2,p_3,\bar{p}_4)+(2 3)+(24)\Big]\\[2ex]
&p_{1\,\nu_1}\,V_{E}^{\m_1\n_1\m_2\n_2\m_3\n_3\mu_4\nu_4}(p_1,p_2,p_3,\bar{p}_4)=\notag\\
&=\Big[4\, \mathcal{B}^{\mu_1\hspace{0.4cm}\mu_2\nu_2\mu_3\nu_3}_{\hspace{0.3cm}\lambda\nu_1}(p_2,p_3)V_{E}^{\lambda\n_1\mu_4\nu_4}(p_1+p_2+p_3,\bar{p}_4)+(34)+ (24)\Big]\notag\\
&\hspace{0.5cm}+\Big[2 \, \mathcal{C}^{\mu_1\hspace{0.4cm}\mu_2\nu_2}_{\hspace{0.3cm}\lambda\nu_1}(p_2)V_{E}^{\lambda\n_1\m_3\n_3\mu_4\nu_4}(p_1+p_2,p_3,\bar{p}_4)+(2 3)+(24)\Big],
\end{align} 
where  $\mathcal{C}$ and $\mathcal{B}$ are given in \eqref{BB} and \eqref{CC}.
These equations can be generalized to the case of $n$-point functions. 

\section{Decomposition of the 4T} 
The analysis in momentum space allows to identify the contributions generated by the breaking of the conformal symmetry, after renormalization, in a direct manner. One specific feature is the prediction of a nonlocal structure for the effective action, in the form of massless exchanges attached to the external lines of a graviton vertex. \\
 For this purpose we will be using the longitudinal transverse L/T decomposition of the correlator presented in \cite{Bzowski:2013sza} for 3-point functions, extending it to the 4T. This procedure has been investigated in detail for 3-point functions in \cite{Coriano:2018bsy} in the context of a perturbative approach \cite{Coriano:2018zdo}. The perturbative analysis in free field theory shows how renormalization acts on the two L/T subspaces, forcing the emergence of a trace in the longitudinal sector.   \\
Due to the constraint imposed by the conformal symmetry (i.e. their CWI's), the correlation functions can be decomposed into a transverse-traceless and a semilocal part. The term semilocal refers to contributions which are obtained from the conservation and trace Ward identities. Of an external off-shell graviton only its spin-2 component will couple to transverse-traceless part. \\
The split of the energy momentum tensor operator in terms of a transverse traceless ($tt$) part and of a longitudinal (local) part \cite{Bzowski:2013sza} is defined in the form
\begin{equation}
T^{\mu_i\nu_i}(p_i)\equiv t^{\mu_i\nu_i}(p_i)+t_{loc}^{\mu_i\nu_i}(p_i),\label{decT}
\end{equation}
with
\begin{align}
\label{loct}
t^{\mu_i\nu_i}(p_i)&=\Pi^{\mu_i\nu_i}_{\alpha_i\beta_i}(p)\,T^{\alpha_i \beta_i}(p_i)\\
t_{loc}^{\mu_i\nu_i}(p_i)&=\Sigma^{\mu_i\nu_i}_{\alpha_i\beta_i}(p)\,T^{\alpha_i \beta_i}(p_i).
\end{align}
We have introduced the transverse-traceless ($\Pi$), transverse-trace $(\tau)$ and 
longitudinal ($\mathcal{I}$) projectors, given respectively by 
\begin{align}
\label{prozero}
\pi^{\mu}_{\alpha} & = \delta^{\mu}_{\alpha} - \frac{p^{\mu} p_{\alpha}}{p^2},  \qquad \tilde{\pi}^{\mu}_{\alpha} =\frac{1}{d-1}\pi^{\mu}_{\alpha} \\\
\Pi^{\mu \nu}_{\alpha \beta} & = \frac{1}{2} \left( \pi^{\mu}_{\alpha} \pi^{\nu}_{\beta} + \pi^{\mu}_{\beta} \pi^{\nu}_{\alpha} \right) - \frac{1}{d - 1} \pi^{\mu \nu}\pi_{\alpha \beta}\label{TTproj}, 
\end{align}
\begin{align}
\mathcal{J}^{\mu\nu}_{\alpha\beta}&=\frac{1}{p^2}p_{\beta}\left( p^{\mu}\delta^{\nu}_{\alpha} +p^{\nu}\delta^{\mu}_{\alpha} -
\frac{p_{\alpha}}{d-1}( \delta^{\mu\nu} +(d-2)\frac{p^\mu p^\nu}{p^2})    \right)\\
\mathcal{I}^{\mu\nu}_{\alpha\beta}&=\frac{1}{2}\left(\mathcal{J}^{\mu\nu}_{\alpha\beta} +\mathcal{J}^{\mu\nu}_{\beta\alpha}\right) \qquad \tau^{\mu\nu}_{\alpha\beta} =\tilde{\pi}^{\mu \nu}\delta_{\alpha \beta}\\
\mathcal{I}^{\mu\nu}_{\alpha}&=\frac{1}{p^2}\left( p^{\mu}\delta^{\nu}_{\alpha} +
p^{\nu}\delta^{\mu}_{\alpha} -
\frac{p_{\alpha}}{d-1}( \delta^{\mu\nu} +(d-2)\frac{p^\mu p^\nu}{p^2}  \right)\label{proone}\\
\mathcal{I}^{\mu\nu}_{\alpha\beta}&=\frac{1}{2}\left(p_\beta \mathcal{I}^{\mu\nu}_{\alpha}
+ p_\alpha \mathcal{I}^{\mu\nu}_{\beta}\right)\label{protwo}
\end{align}
with 
\begin{align}
\delta^{(\mu}_\alpha\delta^{\nu)}_{\beta}&=\Pi^{\mu \nu}_{\alpha \beta} +\Sigma^{\mu\nu}_{\alpha\beta} \\
\Sigma^{\mu_i\nu_i}_{\alpha_i\beta_i}&\equiv\mathcal{I}^{\mu_i\nu_i}_{\alpha_i\beta_i} +\tau^{\mu_i\nu_i}_{\alpha_i\beta_i}\notag\\
&=\frac{1}{p_i^2}\Big[2\delta^{(\nu_i}_{(\alpha_i}p_i^{\mu_i)}p_{i\,\beta_i)}-\frac{p_{i\alpha_i}p_{i\beta_i}}{(d-1)}\left(\delta^{\mu_i\nu_i}+(d-2)\frac{p_i^{\mu_i}p_i^{\nu_i}}{p_i^2}\right)\Big]+\frac{1}{(d-1)}\pi^{\mu_i\nu_i}(p_i)\delta_{\alpha_i\beta_i}\label{Lproj}.
\end{align}
Notice that we have combined together the operators $\mathcal{I}$ and $\tau$ into a projector 
$\Sigma$ which defines the local components of a given tensor $T$, according to \eqref{loct}, which are proportional both to a given momentum $p$ (the longitudinal contribution) and to the trace parts. Both $\Pi$ and $\tau$ are transverse by construction, while $\mathcal{I}$ is longitudinal and of zero trace.

The projectors induce a decomposition respect to a specific momentum $p_i$.
By using \eqref{decT}, the entire correlator is written as
\begin{align}
&\braket{T^{\mu_1\nu_1}(p_1)T^{\mu_2\nu_2}(p_2)T^{\mu_3\nu_3}(p_3)T^{\mu_4\nu_4}(\bar{p}_4)}=\notag\\
&=\braket{t^{\mu_1\nu_1}(p_1)t^{\mu_2\nu_2}(p_2)t^{\mu_3\nu_3}(p_3)t^{\mu_4\nu_4}(\bar{p}_4)}+\braket{T^{\mu_1\nu_1}(p_1)T^{\mu_2\nu_2}(p_2)T^{\mu_3\nu_3}(p_3)T^{\mu_4\nu_4}(\bar{p}_4)}_{loc}\label{decTTTT}
\end{align} 
where the first contribution is the transverse-traceless part which satisfies  by construction the conditions
\begin{equation}
\begin{split}
p_{i\,\mu_i}\braket{t^{\mu_1\nu_1}(p_1)t^{\mu_2\nu_2}(p_2)t^{\mu_3\nu_3}(p_3)t^{\mu_4\nu_4}(\bar{p}_4)}&=0,\qquad i=1,2,3,4\,,\\
\delta_{\mu_i\nu_i}\braket{t^{\mu_1\nu_1}(p_1)t^{\mu_2\nu_2}(p_2)t^{\mu_3\nu_3}(p_3)t^{\mu_4\nu_4}(\bar{p}_4)}&=0,\qquad i=1,2,3,4.
\end{split}\label{ttproperties}
\end{equation}
It is clear now that only the second term in \eqref{decTTTT} contributes entirely to the conservation WIs. Thus, the proper new information on the form factors of the 4-point function is entirely encoded in its transverse-traceless ($tt$) part, since the remaining longitudinal + trace contributions,  corresponding to the local term, are related to lower point functions. \\
The derivation of the anomaly part of the correlator in the 4T has been worked out in \cite{Coriano:2021nvn}. One can show that its structure can be summarized in the form
\begin{align}
&\braket{T^{\mu_1\nu_1}(p_1)T^{\mu_2\nu_2}(p_2)T^{\mu_3\nu_3}(p_3)T^{\mu_4\nu_4}(\bar{p}_4)}^{(d=4)}_{anomaly}=\notag\\
&\qquad=\braket{T^{\mu_1\nu_1}(p_1)T^{\mu_2\nu_2}(p_2)T^{\mu_3\nu_3}(p_3)T^{\mu_4\nu_4}(\bar{p}_4)}_{poles}\notag\\
&\qquad+\braket{T^{\mu_1\nu_1}(p_1)T^{\mu_2\nu_2}(p_2)T^{\mu_3\nu_3}(p_3)T^{\mu_4\nu_4}(\bar{p}_4)}_{0-residue}
\end{align}
where the first contribution is anomalous (Weyl-variant) and the second one
is traceless
\begin{equation}
	\delta_{\mu_i\nu_i}\braket{T^{\mu_1\nu_1}(p_1)T^{\mu_2\nu_2}(p_2)T^{\mu_3\nu_3}(p_3)T^{\mu_4\nu_4}(\bar{p}_4)}_{0-residue}=0,\qquad i=1,2,3,4. 
\end{equation}
We call it the "zero residue" or the "zero trace" (0T) part, since the operation of tracing the anomalous part removes the anomaly pole in the bilinear mixing terms, leaving a residue which is proportional to the anomaly and is given in \cite{Coriano:2021nvn}. This part carries no pole. 
On the other hand, the anomaly part is then explicitly given as
\begin{align}
&\braket{T^{\mu_1\nu_1}(p_1)T^{\mu_2\nu_2}(p_2)T^{\mu_3\nu_3}(p_3)T^{\mu_4\nu_4}(\bar{p}_4)}_{poles}=\notag\\
&=\frac{\pi^{\mu_1\nu_1}(p_1)}{3}\,\braket{T(p_1)T^{\mu_2\nu_2}(p_2)T^{\mu_3\nu_3}(p_3)T^{\mu_4\nu_4}(\bar{p}_4)}_{anomaly}+(perm.)\notag\\
&-\frac{\pi^{\mu_1\nu_1}(p_1)}{3}\frac{\pi^{\mu_2\nu_2}(p_2)}{3}\,\braket{T(p_1)T(p_2)T^{\mu_3\nu_3}(p_3)T^{\mu_4\nu_4}(\bar{p}_4)}_{anomaly}+(perm.)\notag\\
&+\frac{\pi^{\mu_1\nu_1}(p_1)}{3}\frac{\pi^{\mu_2\nu_2}(p_2)}{3}\frac{\pi^{\mu_3\nu_3}(p_2)}{3}\,\braket{T(p_1)T(p_2)T(p_3)T^{\mu_4\nu_4}(\bar{p}_4)}_{anomaly}+(perm.)\notag\\
&-\frac{\pi^{\mu_1\nu_1}(p_1)}{3}\frac{\pi^{\mu_2\nu_2}(p_2)}{3}\frac{\pi^{\mu_3\nu_3}(p_3)}{3}\frac{\pi^{\mu_4\nu_4}(p_4)}{3}\,\braket{T(p_1)T(p_2)T(p_3)T(\bar{p}_4)}_{anomaly}.
\end{align}
\begin{figure}[t]
 \centering
	\raisebox{-1.5ex}{\includegraphics[scale=0.5]{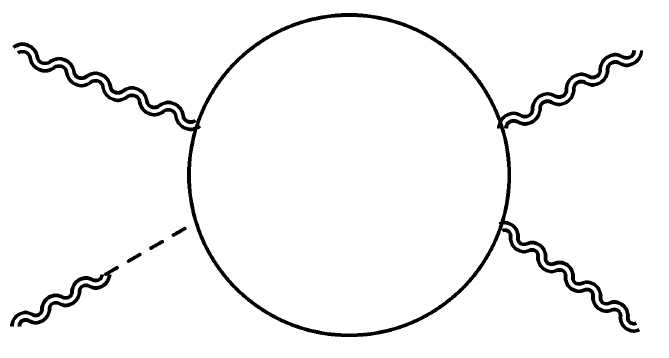}} 
	\raisebox{-1.5ex}{\includegraphics[scale=0.5]{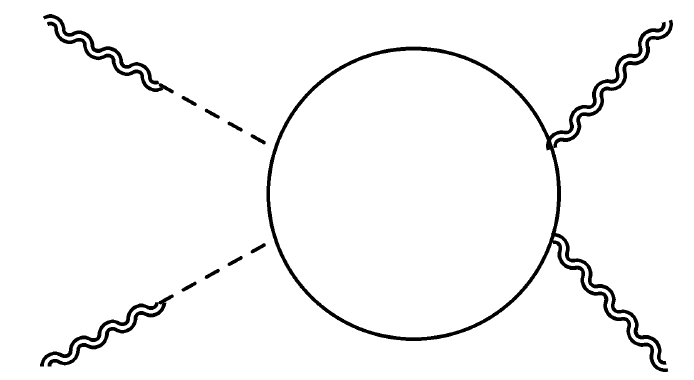}} 
	\raisebox{-1.5ex}{\includegraphics[scale=0.5]{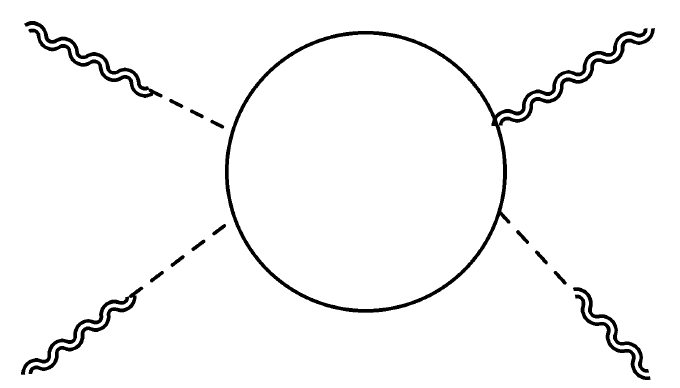}} 
	\raisebox{-1.5ex}{\includegraphics[scale=0.5]{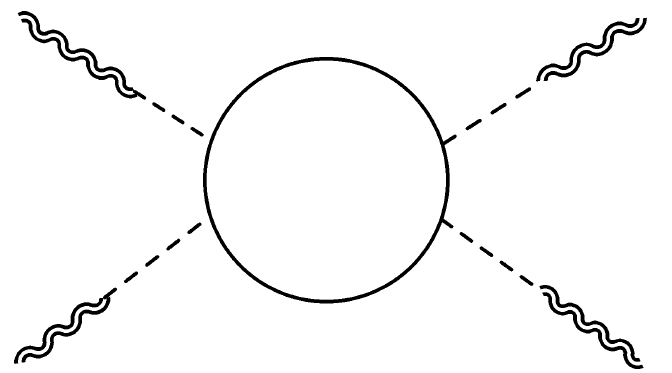}} 
	\caption{The Weyl-variant contributions from $\mathcal{S}_A$  to the renormalized vertex for the 4T with the corresponding bilinear mixings in $d=4$ \label{4T}}
\end{figure}
The result is exemplified in Fig. \eqref{4T}. The dashed lines in this figure correspond to nonlocal exchanges identified by the CWI's in the anomaly part of the correlator in the form of anomaly poles around flat space. In the case of the $TTT$ the pattern is similar and it is reproduced by the  nonlocal action that we will describe next.
Our result for the anomaly action, as predicted by the CWIs, in an expansion in terms of the fluctuations $h$, can then be collected into the form 
\begin{align}
\label{res}
\mathcal{S}_A=&\int d^4 x_1 d^4 x_2 \langle T\cdot h(x_1)T\cdot h(x_2)\rangle +
\int d^4 x_1 d^4 x_2 d^4 x_3 \langle T\cdot h(x_1)T\cdot h(x_2)T\cdot h(x_3)\rangle_{pole}
\notag \\
&+ \int d^4 x_1 d^4 x_2 d^4 x_3 d^4 x_4 
\left( \langle T\cdot h(x_1)T\cdot h(x_2)T\cdot h(x_3)T\cdot h(x_4)\rangle_{pole} +\right.
\notag \\
&\qquad\qquad\qquad\hspace{2cm} \left. + \langle T\cdot h(x_1)T\cdot h(x_2)T\cdot h(x_3)T\cdot h(x_4)\rangle_{0T}\right),
\end{align}
where we have also included the (complete) $TT$, plus the extra traceless (0T) term appearing in the 4T.
 More details can be found in \cite{Coriano:2021nvn}. It is then clear that around Minkowski space there is no sign of a local dilaton interaction, rather, the theory is characterised by the exchange of what appears to be a sequence of anomaly poles. The role of the 
 $0T$ term is a new component that appears at quartic order in the gravitational fluctuations and needs a separate analysis in the context of nonlocal actions, which is underway. Notice that in coordinate space, these contributions to the effective action are equivalent to repeated insertions of the $R \Box^{-1}$ operator. \\

\section{Evading Lovelock's theorem with $V_E$}
The action described in the previous section is well constrained by the CWI's and can be explicitly computed around flat space. Perturbative tests and general solutions of the CWI's provide a consistent picture of an expansion identified by a decomposition of the form given in Fig.\eqref{4T}. Notice that, apart from the Weyl invariant terms contained in the renormalized action $\sm_R(g)$, the rest of this action is entirely fixed by the two conterterms $V_E$ and $V_{C^2}$ introduced in \eqref{ffr}. If we neglect the $C^2$ part and consider only the Gauss Bonnet term, it is clear that the anomaly contribution is, in this specific case, only associated with the presence of a topological term evaluated in $d$ dimensions, accompanied by an extra $1/\epsilon$ factor needed for the regularization of $\sm(g)$. \\
Recently, the possibility of evading Lovelock's theorem at $d=4$, via a singular redefinition of the dimensionless coupling of the Gauss-Bonnet term, has been widely discussed in the area of cosmology and general relativity. The term is added as a quadratic contribution of the curvature tensor to the Einstein-Hilbert action, defining a theory of  "Einstein Gauss-Bonnet" (EGB) type. The procedure shares close similarities with the one that we have discussed above for the conformal anomaly action, since the use of the $1/\epsilon V_E$ counterterm is identical in both types of theories. Obviously, the context in which such EGB theories are defined is purely classical and essentially exploit the fact that a counterterm of the form $1/\epsilon V_E$  correspond to a finite renormalization rather than to an ordinary singular one. This point has been addressed in the recent literature \cite{Coriano:2022ftl}. In this respect, the use of such term in dimensional regularization opens up the way for new forms of dilaton gravities, which can be of cosmological interest, for generating equations of motion of the second order. Such theories are not unique, in a way, since the "regularization" procedure involved in order to induce a nonvanishing topological term at $d=4$ from non integer dimensions, is not unique. We refer to \cite{Coriano:2022ftl} for a discussion of many of the points that we are going to summarize. 

 \section{Towards a nonlocal EGB theory}
As mentioned in our introduction, corrections to the Einstein-Hilbert action, in the form of higher powers of the Riemann curvature $R_{\mu\nu\alpha\beta}$ and of its contractions, have been extensively discussed in the cosmological context. The goal of such modifications is to deepen our insight into the problem of the dark energy dominance in the evolution of our universe, providing an answer to the hierarchy problem, implicit in the current fits of the cosmological constant, within the $\Lambda$CDM ($\Lambda$-Cold Dark Matter) model \cite{Antoniadis:2011ib}. \\
One of the fundamental issue that higher derivative gravity theories have to face, is due to the presence of ghost states in their spectra, at quantum level. 
In general  corrections in the form of terms in the Lagrangian of dimension-4, quadratic in the Riemann tensor, may appear in the form of Weyl-invariant densities, multiplied by dimensionless couplings. Some of these invariants may generate equations of motion of the second order, and are classified by Lovelock's theorem \cite{Lovelock:1971yv} (see also \cite{Charmousis:2014mia}). The theorem guarantees, that the Einstein Hilbert action with a cosmological constant $\Lambda$
\beq
\mathcal{S}_{EH}=\int d^d x \sqrt{g}(M_P^2 R + \Lambda)  
\eeq 
is the only one that provides equations of motion of the second order at $d=4$. Example of topological actions are, for instance, 
the EH action $d=2$, which is metric independent. This action shares similar characteristics to other actions that also include topological terms ($E_4, E_6$ etc) quadratic and cubic in $R_{\mu\nu\alpha\beta}$ as, for example,  in $d=4$ and  6. 
The inclusion of such topological corrections via a singular coupling limit - from $d\neq 4$ to $d=4$ - essentially extends the known 
"regularization" procedure at $d=2$ for the EH term, to the 4-dimensional invariant $\sqrt{g} E$ \cite{Glavan:2019inb}. However, for the procedure to be consistent, it has to be correctly framed in the context of dimensional regularization (DR)\cite{Gurses:2020rxb}, with significant modifications of the original proposal. \\
 In a consistent formulation, the limit generates, in this way, a $0/0$ term in the action which is finite, but that includes also a dilaton in the spectrum  \cite{Hennigar:2020lsl,Fernandes:2020nbq} and is therefore equivalent to a theory of dilaton gravity.
The action takes the counterterm form 
\beq
\label{GB0}
\alpha V_E(d)=\alpha \int d^d x \sqrt{g}E, 
\eeq
where $E\equiv E_4$ is the Euler-Poincar\`e topological density \eqref{fourd} and 
$V_E$, as just mentioned, is added to the Einstein-Hilbert action (EH), 
\beq
\mathcal{S}_{EGB}=S_{EH} + \alpha V_E,
\label{first}
\eeq
while $\alpha$ is a dimensionless coupling constant. The addition of this term, in general,  
generates equations of motion of the form 
 \beq
\frac{1}{\kappa}\left(R_{\mu\nu}-\frac{1}{2}g_{\mu\nu}R+\Lambda_{0}g_{\mu\nu}\right)+\alpha (V_{E}(d))_{\mu\nu}=0,
\label{GB2}
\eeq
with $(V_{E}(d))_{\mu\nu}={\delta V_E(d)}/{\delta g_{\mu\nu}}$.
However, due to the topological nature of \eqref{GB0} at $d=4$, the variation ($\delta_g$) of the integrand 
in \eqref{GB0} with respect to the metric, is an ordinary boundary contribution
\begin{equation}
\delta_g (\sqrt{g} E)=\sqrt{g}\,\nabla_\sigma\, \delta_g X^\sigma,\quad
\delta_g X^\sigma=\varepsilon^{\mu\nu\alpha \beta}\varepsilon^{\sigma\lambda\gamma\tau}
\delta_g \Gamma^\eta_{\nu\lambda}g_{\mu\eta}R_{\alpha \beta \gamma \tau}, 
\end{equation}
where $\varepsilon^{\mu\nu\alpha \beta}={\epsilon^{\mu\nu\alpha \beta}}/{\sqrt{g}}$, and $\Gamma$ is the Christoffel connection in the metric $g_{\mu\nu}$.  
The term $(V_E(d))^{\mu\nu}$ is therefore an evanescent term at $d=4$
\begin{equation}
(V_E(4))^{\mu\nu}= 4R_{\mu\alpha\beta\sigma}R^{\;\,\alpha\beta\sigma}_\nu-8R_{\mu\alpha\nu\beta}R^{\alpha\beta}-8R_{\mu\alpha}R^{\;\,\alpha}_{\nu}+4RR_{\mu\nu}-g_{\mu\nu}{E}=0,    \qquad (d=4)\nn\\
\end{equation}
if we assume asymptotic flatness and strictly $d=4$. \\
In performing the $d\to 4$ limit, the extra components of the metric appear in $d=4$ as in any ordinary compactification of metrics with extra dimensions. The natural approach is to re-express the metric in terms of a fiducial metric and of a dilaton factor as in \eqref{cf1}.\\ 
The result is the derivation of a quartic dilaton gravity that differs from the expression of 
$\sm(g)$ by a Weyl-invariant part coming from the quantum corrections and by the inclusion of two - rather than one - counterterms.
Here we are going to provide a summary of these developments, reviewing such recent proposal  and considering the possibility of performing an {\em additional} finite renormalization of such term $V_E$, in order to regulate such theories in a nonlocal form.  
For related discussion of the local theory we refer to\cite{Fernandes:2020nbq,Hennigar:2020lsl,Lu:2020iav}.
We remark that the evanescent character of $V_E(d)$ at $d=4$ is what renders this procedure interesting. As already pointed out, the Gauss-Bonnet term is necessary in order for the effective action to satisfy the WZ consistency condition \eqref{WZ} but corresponds to a finite renormalization of the effective action rather than to an infinite one: the singularities of the loop corrections in $\sm(g)$ are canceled only by the inclusion of the Weyl term $V_{C^2}$.

\section{Conformal decompositions and Wess-Zumino actions}
The role of $V_E$ and its extensions can be investigated by enlarging the integration region of $V_E$ from 4 to $d$ dimensions, as in \eqref{GB0}, and the corresponding limiting action 

\beq
\label{bonc}
\mathcal{S}_{EGB} =\mathcal{S}_{EH} +\mathcal{S}_{GB}(d) \qquad \mathcal{S}_{GB}(d)=\frac{\alpha}{\epsilon}V_E(d)
\eeq
naively appears to be finite at $d=4$, and modifies \eqref{GB0}. If we set aside the problem of consistency of the $0/0$ limit, the theory that results appears to be purely gravitational and quadratic in $R_{\mu\nu\rho\sigma}$. The claim of the existence of a purely gravitation theory ghost-free, beyond Einstein's formulation, is based on this interesting observations.  \\  
Such singular limit seem to provide a theory that violates Lovelock's theorem. We recall that Lovelock's theorem states that at $d=4$ only the EH action plus a cosmological constant generate equations of motion of second order involving only the metric.  
However, as already mentioned, the tensorial limit of the equations is not well defined \cite{Gurses:2020rxb}. In fact, $V_E^{\mu\nu}$ computed around flat space, once it is extrapolated from generic $d$ to $d=4$ \cite{Fernandes:2020nbq,Hennigar:2020lsl,Lu:2020iav}, does not exhibit a compensating factor $O(\epsilon)$, which is necessary in order to erase the singular behaviour of the coupling. One needs a more general approach in order to define a consistent theory. \\
A modified procedure that can be applied to this case, well-known in the case of the $d=2$ EH action, which is also topological, is to regulate the singular action by a method based on the Weyl-gauging of the metric. The symmetry \eqref{cf2}  is going to be broken after we complete the regularization of the new action \cite{Coriano:2022ftl}. The outcome is the emergence of  a constraint between the equations of motion of $\bar{g}_{\mu\nu}$ and $\phi$. The appearance of such constraint is not surprising, since its origin is, in a way, anomaly related. In fact, as we are going to show below, the regulated action is an ordinary WZ action, and the constraint is generated quite naturally by acting with a conformal variation on its finite expression. We proceed by illustrating this point in more detail. \\
 The WZ action of the GB term $V_E(g_{\mu\nu}, d)$ is defined as the difference 
\beq
\mathcal{S}^{(WZ)}_{GB}\equiv\frac{\alpha}{\epsilon}\left(V_E(\bar{g}_{\mu\nu}e^{2\phi},d)- V_E(\bar{g}_{\mu\nu},d)\right)
\label{ss}
\eeq
where 
\beq
\frac{1}{\epsilon}\frac{\delta V_E(\bar{g}_{\mu\nu}e^{2\phi},d)}{\delta\phi}=\sqrt{g}E(g)
\eeq
while the expansion in $\epsilon$ around $d=4$ for $V_E(\bar{g}_{\mu\nu}e^{2\phi},d)$
\beqa 
\frac{1}{\epsilon}V_E(\bar{g}_{\mu\nu}e^{2\phi},d)&=&\frac{1}{\epsilon}V_E(\bar{g}_{\mu\nu},d=4) + 
V'_E(\bar{g}_{\mu\nu},\phi, d=4)\nonumber \\
\label{exp}
\eeqa
and a similar one for  $V_E(\bar{g}_{\mu\nu},d)$, in the limit $\epsilon\to 0$,  combined 
give 
\beq
\mathcal{S}^{(WZ)}_{GB}=\alpha V'_E(\bar{g}_{\mu\nu},\phi, d=4), 
\label{ww}
\eeq
having dropped a Weyl-invariant terms $V'_E(\bar{g}_{\mu\nu},d=4)$ in the dimensional expansion around $d=4$. The dimensional reduction that generates \eqref{ww} requires some care, due to the presence of cutoffs in the extra dimensions. They identify Weyl-invariant terms that are dropped while performing the regularization, and are discussed in 
\cite{Coriano:2022ftl}. 

Neglecting some Weyl-invariant terms, $V'_E$ 
can be given in the form derived in \cite{Fernandes:2020nbq,Hennigar:2020lsl,Lu:2020iav} 
\begin{equation}
\mathcal{S}^{(WZ)}_{GB}=\alpha\int d^4 x \sqrt{\bar{g}} \biggl[\phi \bar{E}-\Big(4 \bar{G}^{\mu\nu}(\bar{\nabla}_\mu\phi\bar{\nabla}_\nu\phi)+2(\bar{\nabla}_\lambda \phi\bar{\nabla}^\lambda \phi)^2
 +4\bar\Box\phi\bar{\nabla}_\lambda \phi\bar{\nabla}^\lambda \phi\Big)\biggl],
 \label{GGB}
\end{equation}
where $\bar{G}_{\mu\nu}$ is the Einstein tensor in the fiducial metric $\bar{g}_{\mu\nu}$. Counterexamples to such  procedure, where the action is not properly regulated, are those in which the $d-4$ extra space components of $\bar{g}_{\mu\nu}$ are flat, as discussed in \cite{Gurses:2020rxb}. Such cases need to be excluded, for consistency. Notice that the WZ action is clearly dependent separately on $\bar{g}$ and $\phi$, and this explains why the original invariance under the conformal decomposition \eqref{cf2} is broken.  
The equations of motion for the metric, from   
$\bar{g}_{\mu\nu}$ and $\phi$ are constrained by the relations 
\beqa
\label{fin}
\left(2 {g}_{\mu\nu}\frac{\delta}{\delta _{\mu\nu}}-\frac{\delta}{\delta\phi}\right)\mathcal{S}^{(WZ)}_{GB}=-\alpha\sqrt{\bar g}\bar{E},
\eeqa
\beqa
&&2{g}_{\mu\nu}\frac{\delta \mathcal{S}^{(WZ)}_{GB}}{\delta {g}_{\mu\nu}}=\sqrt{g}E - 
\sqrt{\bar g}\bar E, 
\label{equal}
\eeqa
\beq
\frac{\delta \mathcal{S}^{(WZ)}_{GB}}{\delta \phi}=\sqrt{g}E.
\eeq
One may recognize the similarity between these relations and those typical of the WZ actions derived from the analysis of the conformal anomaly. The difference between those theories and the current case lays in the absence of Weyl invariant terms in the EGB, 
which are, instead, naturally present in the case in which the same action is obtained as a conformal backreaction. Beside, the $C^2$ term is clearly absent in a EGB theory.

\section{Conformal Symmetry breaking}
WZ actions have been identified in the past either by the Weyl gauging procedure, as shown above, or, equivalently, by the Noether method. Both methods stop at order $d$ in the dilaton field, if we are in a $d$-dimensional spacetime, and generate anomaly actions which are not purely gravitational.  The actions obtained are local, since they include the dilaton field. If we were able to solve the equations 
of motion for $\phi$, then the actions would become nonlocal, in agreement with the general result that anomalies are not related to local actions. \\
One immediately realizes that the separation of the conformal factor from the fiducial 
metric has necessarily to be associated with the breaking of the conformal symmetry.  
  In these formulations, the dilaton $\phi$ can either be part of the regulated action, as discussed in \cite{Fernandes:2020nbq,Hennigar:2020lsl,Lu:2020iav}, or can be eliminated, by solving for this field in terms of the background metric.\\
   In the Lagrangians of the first class 
 (dilaton gravities), $\phi$ is generally interpreted as a physical particle. In this case it is convenient to introduce the parameterization 
 \beq
 e^\phi=1- \frac{\chi}{f}
 \eeq
 where $f$ is a conformal breaking scale,
 which breaks the symmetry in both the first and the second term of \eqref{ss}. The action can be expanded around the field value $\phi=\phi_0=0$ (i.e. $ \chi=\chi_0=0$), but it is quite obvious that the selection of such values for $\phi$ or, equivalently,  $\chi$, should be induced by an extra potential which is clearly not part of in $S^{(GB)}_{WZ}$. A mechanism of spontaneous breaking of the local conformal symmetry needs to be invoked in order to stabilize the dilaton.
 Notice that $S_{WZ}$ already breaks the symmetry \eqref{ss} by the anomaly since 
 \beq
 \frac{\delta S^{(GB)}_{WZ}}{\delta \sigma}=-\sqrt{\bar g}\bar{E}. 
 \eeq
  Actions belonging to this class, quartic in the field $\phi$, obviously depend on the choice of the form of 
 $\bar{g}_{\mu\nu}$ in DR and  may differ, therefore,  by finite $\phi$-dependent terms. \\
 This is well-known in the case of ordinary Kaluza-Klein theories, whose Lagrangians depend on the manifold of compactification. 
A discussion of these issues has been presented in \cite{Coriano:2022ftl}. Actions in which the dilaton cannot be eliminated are suitable for a description of the spontaneous breaking of the conformal symmetry. The second class of actions, where the dilaton is removed, exhibit 
the $R\Box^{-1}$ terms that we have identified in the previous sections from the analysis of the conformal Ward identities satisfied by the quantum effective action.

As we are going to show, also in the EGB case, a specific finite renormalization of $V_E$ generates an action which is only quadratic in $\phi$, and may capture consistently the UV behaviour of the anomaly. It can be written in a nonlocal form. In this case the expansion is in terms of $R\Box^{-1}$, as discussed in  \cite{Coriano:2021nvn} \cite{Coriano:2017mux}. Also in this case,  
the procedure does not allow to identify some Weyl-invariant contributions which are only part of an exact computation of the functional expansion. Direct computations of 3- and 4-point  functions, in specific backgrounds, from flat to Weyl-flat, are the only possibility in order 
to work out the complete expression of such actions.

\section{Nonlocal EGB theories}
If EGB theories can be formulated as dilaton gravities, quite close to ordinary anomaly actions, nonlocal versions of such actions are also possible. Indeed, a modification of the GB term at $O(\epsilon)$, generates anomaly actions quite different from those presneted in \cite{Fernandes:2020nbq,Hennigar:2020lsl,Lu:2020iav}. We have the possibility of performing a finite renormalization of the GB term in DR, as already discussed in \cite{Mazur:2001aa} in the context of conformal anomaly actions, to end up with a very different theory.

For this reason, one can simply modify by a finite  renormalization  the structure of the GB term away from $d=4$ in the form
 \beq
E_{ext}=E + \frac{\epsilon}{2 (d-1)^2} R^2,\qquad \tilde{V}_E=\int d^d x \sqrt{g}E_{ext},
 \eeq
 with $V_E\to \tilde{V}_E$ in \eqref{ss}, obtaining

\begin{equation}
\mathcal{S}^{(WZ)}_{GB} =\frac{\alpha}{\epsilon}\left(\tilde{V}_E(\bar{g}_{\mu\nu}e^{2\phi},d)- \tilde{V}_E(\bar{g}_{\mu\nu},d\right).
\label{inter}
\eeq
At this stage we perform the $d\to 4$ limit and an expansion similar to \eqref{exp}, to derive the regulated GB action
\begin{equation}
\mathcal{S}^{(WZ)}_{GB} = \alpha\int\,d^4x\,\sqrt{-\bar g}\,\left\{\left(\overline E_4 - {2\over 3}
\bar{\Box} \overline R\right)\phi + 2\,\phi\bar\Delta_4\phi\right\},\,
\label{WZ2}
\end{equation}
where $\bar{\Delta}_4$ is the quartic conformally covariant operator 
\cite{Riegert:1987kt}
 \begin{equation}
\Delta_4 \equiv \Box^2 + 2 R^{\mu\nu}\nabla_\mu\nabla_\nu +{1\over 3} (\nabla^\mu R)
\nabla_\mu - {2\over 3} R \Box,
\label{Delta}
\end{equation}
evaluated with respect to the fiducial metric $\bar{g}_{\mu\nu}$. \\
Clearly, Eq. \eqref{WZ2} shows that it is possible to define a quadratic - and not only a quartic action in $\phi$, as before, for the regularized GB term. If add the resulting action to the EH action, we end up with a version of the EGB theory which is purely gravitational, since $\phi$ can be eliminated.  In order to to derive such nonlocal expression, we can use the relation
\beq
\frac{\delta}{\delta\phi}\frac{1}{\epsilon}\tilde{V}_E(g_{\mu\nu},d)= \sqrt{g}\left(E-\frac{2}{3}\Box R +
\epsilon\frac{R^2}{2(d-1)^2}\right)
\eeq
in \eqref{inter}, giving  
 \beqa
 \frac{\delta \mathcal{S}^{(WZ)}_{GB}}{\delta\phi}&=&\alpha\sqrt{g}\left(E-\frac{2}{3}\Box R \right)\nonumber \\
&=&\alpha\sqrt{\bar g}\left(\bar E-\frac{2}{3}\bar \Box\bar R + 4 \bar\Delta_4 \phi\right).
\label{solve}
\eeqa
Using the relation $\sqrt{\bar{g}}\,\bar{\Delta}_4= \sqrt{g}\,\Delta_4 $, valid on conformal scalars, 
and introducing the Green function $D_4(x,y)$ of $\Delta_4$
\beq
\sqrt{g}\,\Delta_4 D_4(x,y)=\delta^4(x,y),
\eeq
we can solve for $\phi$ in \eqref{solve}. Inserting its expression back into \eqref{WZ2} 
we obtain the nonlocal action
\beqa
 \mathcal{S}^{(WZ)}_{GB}& =& 
{\alpha\over 8} \int d^4x\,\sqrt{-g}\, \int d^4x'\,\sqrt{-g'}\,
\left(E_4 - {2\over 3} \Box R\right)_x\, \nonumber \\
&&\qquad \times D_4(x,x')\left(E_4- {2\over 3} \Box R\right)_{x'},\,
\label{anomact}
\eeqa
that coincides with the result provided in \cite{Mazur:2001aa} by Mazur and Mottola. \\
  As clear from \eqref{solve}, we have removed the dilaton $\phi$ by solving that equation by the Green function of $\Delta_4$. Athough 
  the operator is quartic, however, as discussed in recent analysis, the functional expansion of nonlocal actions of this type in the background metric can be organized in terms of operatorial insertions of the form $R\Box^{-1}$ for each external leg, at least around flat space \cite{Coriano:2017mux,Coriano:2021nvn}. This is in agreement with the discussion introduced in the previous sections, at least up to 3-point functions (TTT). 

\section{ Conclusions}
We have reviewed recent work on the relation between traditional anomaly actions, in the form of dilaton gravity theories, and compared them against recent proposals of EGB theories, introduced by an infinite regularization of the coupling of the topological Gauss-Bonnet term. Topological terms in gravity, introduced in a purely classical context, share a significant overlap with former analysis of the anomaly actions of Wess-Zumino forms, widely discussed in the related literature. This has sparked a wide interest in the General Relativity community, for providing a way to bypass  Lovelock's theorem, that clearly is not framed in a context of dimensional regularization. As we have seen, a finite renormalization of the GB  density, which is perfectly allowed in DR, gives the possibility of generating an EGB theory which is nonlocal. A final comment concerns the important role that these class of actions play in the study of topological materials through effecctive field theory descriptions \cite{Chernodub:2019tsx,Chernodub:2021nff}. Work in this direction is underway. 

\centerline{\bf Acknowledgements}
The work of M.C. is supported by a PON fellowship on "Topological Insulators" and partially supported by the Italian Institute of Technology. The work of C.C. S.L. and R.T. is supported by INFN iniziativa specifica QFT-HEP. M. M. M. is supported by the European Research Council (ERC) under the European Union as Horizon 2020 research and innovation
program (grant agreement No818066) and by Deutsche Forschungsgemeinschaft (DFG, German Research Foundation) under Germany's Excellence Strategy EXC-2181/1 - 390900948 (the Heidelberg
STRUCTURES Cluster of Excellence).
We dedicate this work to the memory of our friend and collaborator Theodore Tomaras who will be deeply missed. 

 \providecommand{\href}[2]{#2}\begingroup\raggedright\endgroup


\begin{thebibliography}{10}

\bibitem{Visser:2002ew}
M.~Visser, {\it {Sakharov's induced gravity: A Modern perspective}},  {\em Mod.
  Phys. Lett. A} {\bf 17} (2002) 977--992,
  [\href{http://xxx.lanl.gov/abs/gr-qc/0204062}{{\tt gr-qc/0204062}}].

\bibitem{1984PhLB..134...56R}
R.~J. {Riegert}, {\it {A non-local action for the trace anomaly}},  {\em
  Physics Letters B} {\bf 134} (Jan., 1984) 56--60.

\bibitem{Bzowski:2013sza}
A.~Bzowski, P.~McFadden, and K.~Skenderis, {\it {Implications of conformal
  invariance in momentum space}},  {\em JHEP} {\bf 03} (2014) 111,
  [\href{http://xxx.lanl.gov/abs/1304.7760}{{\tt arXiv:1304.7760}}].

\bibitem{Coriano:2018bsy}
C.~Corian\`o and M.~M. Maglio, {\it {The general 3-graviton vertex ($TTT$) of
  conformal field theories in momentum space in $d =4$}},  {\em Nucl. Phys.}
  {\bf B937} (2018) 56--134, [\href{http://xxx.lanl.gov/abs/1808.1022}{{\tt
  arXiv:1808.1022}}].

\bibitem{Coriano:2018bbe}
C.~Corian\`o and M.~M. Maglio, {\it {Exact Correlators from Conformal Ward
  Identities in Momentum Space and the Perturbative $TJJ$ Vertex}},  {\em Nucl.
  Phys.} {\bf B938} (2019) 440--522,
  [\href{http://xxx.lanl.gov/abs/1802.0767}{{\tt arXiv:1802.0767}}].

\bibitem{Coriano:2017mux}
C.~Corian\`o, M.~M. Maglio, and E.~Mottola, {\it {TTT in CFT: Trace Identities
  and the Conformal Anomaly Effective Action}},  {\em Nucl. Phys.} {\bf B942}
  (2019) 303--328, [\href{http://xxx.lanl.gov/abs/1703.0886}{{\tt
  arXiv:1703.0886}}].

\bibitem{Giannotti:2008cv}
M.~Giannotti and E.~Mottola, {\it {The Trace Anomaly and Massless Scalar
  Degrees of Freedom in Gravity}},  {\em Phys. Rev.} {\bf D79} (2009) 045014,
  [\href{http://xxx.lanl.gov/abs/0812.0351}{{\tt arXiv:0812.0351}}].

\bibitem{Armillis:2009pq}
R.~Armillis, C.~Corian\`{o}, and L.~Delle~Rose, {\it {Conformal Anomalies and
  the Gravitational Effective Action: The $TJJ$ Correlator for a Dirac
  Fermion}},  {\em Phys. Rev.} {\bf D81} (2010) 085001,
  [\href{http://xxx.lanl.gov/abs/0910.3381}{{\tt arXiv:0910.3381}}].

\bibitem{Armillis:2010qk}
R.~Armillis, C.~Corian\`o, and L.~Delle~Rose, {\it {Trace Anomaly, Massless
  Scalars and the Gravitational Coupling of QCD}},  {\em Phys. Rev.} {\bf D82}
  (2010) 064023, [\href{http://xxx.lanl.gov/abs/1005.4173}{{\tt
  arXiv:1005.4173}}].

\bibitem{Coriano:2018zdo}
C.~Corian\`o and M.~M. Maglio, {\it {Renormalization, Conformal Ward Identities
  and the Origin of a Conformal Anomaly Pole}},  {\em Phys. Lett.} {\bf B781}
  (2018) 283--289, [\href{http://xxx.lanl.gov/abs/1802.0150}{{\tt
  arXiv:1802.0150}}].

\bibitem{Coriano:2021nvn}
C.~Corian\`o, M.~M. Maglio, and D.~Theofilopoulos, {\it {The Conformal Anomaly
  Action to Fourth Order (4T) in $d=4$ in Momentum Space}},
  \href{http://xxx.lanl.gov/abs/2103.1395}{{\tt arXiv:2103.1395}}.

\bibitem{Coriano:2022ftl}
C.~Corian\`o, M.~M. Maglio, and D.~Theofilopoulos, {\it {Topological
  Corrections and Conformal Backreaction in the Einstein Gauss-Bonnet/Weyl
  Theories of Gravity at D=4}},  \href{http://xxx.lanl.gov/abs/2203.0421}{{\tt
  arXiv:2203.0421}}.

\bibitem{Antoniadis:2011ib}
I.~Antoniadis, P.~O. Mazur, and E.~Mottola, {\it {Conformal Invariance, Dark
  Energy, and CMB Non-Gaussianity}},  {\em JCAP} {\bf 1209} (2012) 024,
  [\href{http://xxx.lanl.gov/abs/1103.4164}{{\tt arXiv:1103.4164}}].

\bibitem{Lovelock:1971yv}
D.~Lovelock, {\it {The Einstein tensor and its generalizations}},  {\em J.
  Math. Phys.} {\bf 12} (1971) 498--501.

\bibitem{Charmousis:2014mia}
C.~Charmousis, {\it {From Lovelock to Horndeski`s Generalized Scalar Tensor
  Theory}},  {\em Lect. Notes Phys.} {\bf 892} (2015) 25--56,
  [\href{http://xxx.lanl.gov/abs/1405.1612}{{\tt arXiv:1405.1612}}].

\bibitem{Glavan:2019inb}
D.~Glavan and C.~Lin, {\it {Einstein-Gauss-Bonnet Gravity in Four-Dimensional
  Spacetime}},  {\em Phys. Rev. Lett.} {\bf 124} (2020), no.~8 081301,
  [\href{http://xxx.lanl.gov/abs/1905.0360}{{\tt arXiv:1905.0360}}].

\bibitem{Gurses:2020rxb}
M.~Gurses, T.~C. {S}i{s}man, and B.~Tekin, {\it {Comment on
  ''Einstein-Gauss-Bonnet Gravity in 4-Dimensional Space-Time''}},  {\em Phys.
  Rev. Lett.} {\bf 125} (2020), no.~14 149001,
  [\href{http://xxx.lanl.gov/abs/2009.1350}{{\tt arXiv:2009.1350}}].

\bibitem{Hennigar:2020lsl}
R.~A. Hennigar, D.~Kubiz\v{n}\'ak, R.~B. Mann, and C.~Pollack, {\it {On taking
  the D to 4 limit of Gauss-Bonnet gravity: theory and solutions}},  {\em JHEP}
  {\bf 07} (2020) 027, [\href{http://xxx.lanl.gov/abs/2004.0947}{{\tt
  arXiv:2004.0947}}].

\bibitem{Fernandes:2020nbq}
P.~G.~S. Fernandes, P.~Carrilho, T.~Clifton, and D.~J. Mulryne, {\it
  {Derivation of Regularized Field Equations for the Einstein-Gauss-Bonnet
  Theory in Four Dimensions}},  {\em Phys. Rev. D} {\bf 102} (2020), no.~2
  024025, [\href{http://xxx.lanl.gov/abs/2004.0836}{{\tt arXiv:2004.0836}}].

\bibitem{Lu:2020iav}
H.~Lu and Y.~Pang, {\it {Horndeski gravity as $D \rightarrow 4$ limit of
  Gauss-Bonnet}},  {\em Phys. Lett. B} {\bf 809} (2020) 135717,
  [\href{http://xxx.lanl.gov/abs/2003.1155}{{\tt arXiv:2003.1155}}].

\bibitem{Mazur:2001aa}
P.~O. Mazur and E.~Mottola, {\it {Weyl cohomology and the effective action for
  conformal anomalies}},  {\em Phys.Rev.} {\bf D64} (2001) 104022,
  [\href{http://xxx.lanl.gov/abs/hep-th/0106151}{{\tt hep-th/0106151}}].

\bibitem{Riegert:1987kt}
R.~J. Riegert, {\it {A non-local action for the trace anomaly}},  {\em
  Phys.Lett.} {\bf B134} (1984) 56--60.

\bibitem{Chernodub:2019tsx}
M.~N. Chernodub, C.~Corian\`o, and M.~M. Maglio, {\it {Anomalous Gravitational
  TTT Vertex, Temperature Inhomogeneity, and Pressure Anisotropy}},  {\em Phys.
  Lett.} {\bf B802} (2020) 135236,
  [\href{http://xxx.lanl.gov/abs/1910.1372}{{\tt arXiv:1910.1372}}].

\bibitem{Chernodub:2021nff}
M.~N. Chernodub, Y.~Ferreiros, A.~G. Grushin, K.~Landsteiner, and M.~A.~H.
  Vozmediano, {\it {Thermal transport, geometry, and anomalies}},
  \href{http://xxx.lanl.gov/abs/2110.0547}{{\tt arXiv:2110.0547}}.

\end{thebibliography}
\end{document}